\documentclass[pdftex,amsmath,amssymb,aps,twocolumn,prx]{revtex4-2}
\usepackage[utf8]{inputenc}
\usepackage[T1]{fontenc}
\usepackage{color}
\usepackage{graphicx}
\usepackage{bm}
\usepackage{xr-hyper}
\usepackage{hyperref}
\usepackage{mathtools}

\makeatletter
\newcommand*{\addFileDependency}[1]{
  \typeout{(#1)}
  \@addtofilelist{#1}
  \IfFileExists{#1}{}{\typeout{No file #1.}}
}
\makeatother

\newcommand*{\myexternaldocument}[1]{%
    \externaldocument{#1}%
    \addFileDependency{#1.tex}%
    \addFileDependency{#1.aux}%
}

\myexternaldocument{SI}

\begin{document}

\title{Controlling curvature of self-assembling surfaces via patchy particle design}
\author{Andraž Gnidovec}
\affiliation{
 Faculty of Mathematics and Physics, University of Ljubljana, Ljubljana, Slovenia
}%

\author{Simon Čopar}
\affiliation{
 Faculty of Mathematics and Physics, University of Ljubljana, Ljubljana, Slovenia
}%

\date{\today}

\begin{abstract}
Curved structures in soft matter and biological systems commonly emerge as a result of self-assembly processes where building blocks aggregate in a controlled manner, giving rise to specific system structure and properties. Learning how to precisely tune the curved geometry of these assemblies can in turn elucidate new ways of controlling their functionality. We discuss how one can target self-assembly into surfaces with specified Gaussian curvature in a one-component system of model patchy particles. Given the vast design space of potential patch distributions, we address the problem using an inverse design approach based on automatic differentiation and develop an optimization scheme which solves the exploding gradients problem that arises when we differentiate through long molecular dynamics trajectories. We discuss the model requirements for successful optimization, determine the significant hyperparameter choices influencing algorithm performance and, finally, we demonstrate that we can consistently design patch patterns for assembly into clusters with different target curvature radii.
\end{abstract}

\maketitle

\section{Introduction}

The formation of complex structures in biological systems is often a consequence of spontaneous aggregation of simple building blocks into functional assemblies \cite{Philp1996}. The same principle can be applied in material science where particles are synthesized with specific interactions that direct the self-assembly process towards the target structure, allowing for precise tuning of material properties. Recent advancements in particle synthesis, particularly in designing anisotropic shapes and interactions have significantly broadened the range of obtainable self-assembled structures~\cite{Glotzer2007}. Patchy particles, characterized by highly directional interactions, are particularly well-suited for tackling structural design challenges due to their exceptional flexibility in tuning both the arrangement of patches and the interaction strength between them~\cite{Duguet2016,Gong2017,Moud2022}. Patchiness can arise from charge heterogeneities~\cite{Bianchi2017}, DNA hybridization~\cite{Zhang2021}, or chemical patterning of their surface~\cite{Walther2013}. Adjusting patch properties and their distribution enables the formation of a diverse range of self-assembled structures, including fibers and rings~\cite{Oh2019}, crystalline phases~\cite{Noya2014,Morphew2018,Eslami2024}, as well as other, more complex structures~\cite{Zhang2004}. 

The extensive structural variability and a large interaction design space make it challenging to identify the optimal patch configuration that stabilizes self-assembled materials with target properties. This problem is addressed by inverse design approaches that systematically explore the design space to direct the search towards the most promising candidates~\cite{Sherman2020,Dijkstra2021}. Inverse design algorithms have been effectively applied to colloidal particle self-assembly~\cite{Kumar2019,Sherman2020,Goodrich2021} and have recently demonstrated success in optimizing patchy interactions to assemble desired structures~\cite{Long2018,Romano2020,Lieu2022,Rivera-Rivera2023,King2024}. One of the more general optimization approaches, leveraged across various physics domains, utilizes automatic differentiation (AD) to directly compute derivatives with respect to any parameter in arbitrarily complex simulations \cite{Minkov2020, Thaler2021, Engel2023,Curatolo2023}. In self-assembly processes, AD gives access to kinetic information of the system, which reduces the data requirements compared to data-driven and statistics-based methods~\cite{Miskin2016,Long2018} and allows for optimization of both structural and kinetic properties that are beyond the reach of other inverse design approaches \cite{Goodrich2021,King2024}.

While self-assembly optimization typically focuses on refining local particle interactions, the resulting structures can also be shaped by larger-scale constraints, with system curvature being a particularly significant factor in structure formation. Assembly of mesoscopic particles, such as synthetic building blocks, proteins, or larger biological components, is often influenced by the curvature of the substrate on which assembly occurs~\cite{Xue2022,Mim2012,Shi2023}. The particles themselves can also modify this curvature~\cite{Roux2021} or assemble into curved structures on their own. Hollow, closed structures are of particular interest due to their potential applications as cargo containers for transport~\cite{Naskalska2021}. Examples include protein cages, such as viral capsids and their artificial analogs, where changing the structure of domain proteins enables precise control over the assembled shapes~\cite{Olson2007,Matsuura2018,Laniado2021}. Additionally, spherical nanocapsules have been shown to self-assemble from various anisotropically interacting building blocks, such as coiled-coil peptide modules~\cite{Fletcher2013}, soft colloidal particles~\cite{Evers2016}, and polypeptide–oligonucleotide conjugates~\cite{Wang2021}. Advancing our understanding of how to manipulate the curved geometry of such assemblies can provide new ways to tailor the properties of functional soft materials.

In this paper, we demonstrate how to guide the 3D self-assembly of patchy particles towards the formation of surfaces with a desired Gaussian curvature. We present an optimization scheme based on automatic differentiation that efficiently handles large systems, addressing the challenges of optimizing over chaotic dynamics in molecular dynamics (MD) trajectories~\cite{Metz2022}, which have previously limited investigations to small-scale models~\cite{Goodrich2021,King2024}. We focus on the non-addressable self-assembly limit, i.e. using only a single particle species, with minimal assumptions on the interaction directionality itself and construct a simple but general model for patchy interaction that satisfies these constraints. We further discuss the limits on hyperparameter values where successful optimization is possible in our framework and, finally, we show that we can design patch distributions that lead to particle assemblies with different target radii.

\section{Results}

\subsection{Optimization scheme}

The general computational scheme for AD-based optimization in self-assembly relies on performing MD simulations to model the assembly process, followed by differentiation through the entire MD trajectory to compute derivatives of a given loss function with respect to the interaction parameters \cite{Goodrich2021,King2024}. This approach enables structural design for general features or properties that are difficult to map directly to particle probability distributions -- as required, for example, in optimization methods based on the Kullback-Leibler divergence~\cite{Jadrich2017}. Surface curvature can be considered a general structural feature as targeting a specific local particle environment is too restrictive; particle arrangements in or on curved surfaces usually need to accommodate topological defects, even beyond the minimal required number, as seen in dislocation scars and grain boundaries \cite{Bowick2009}. Instead of constraining local particle arrangements, we consider the overall shape of the structure (i.e. cluster) as the design objective. Our investigation focuses on the case of a single particle species, mimicking the limited number of distinct building blocks that form biological curved surfaces. We note that assemblies with a larger number of different particles usually show better assembly yields \cite{Hubl2024} and the optimal variety of building blocks can be determined using combinatorial design strategies~\cite{Russo2022,Bohlin2023}. 

\begin{figure*}[!ht]
\centering  
\includegraphics[width=\textwidth]{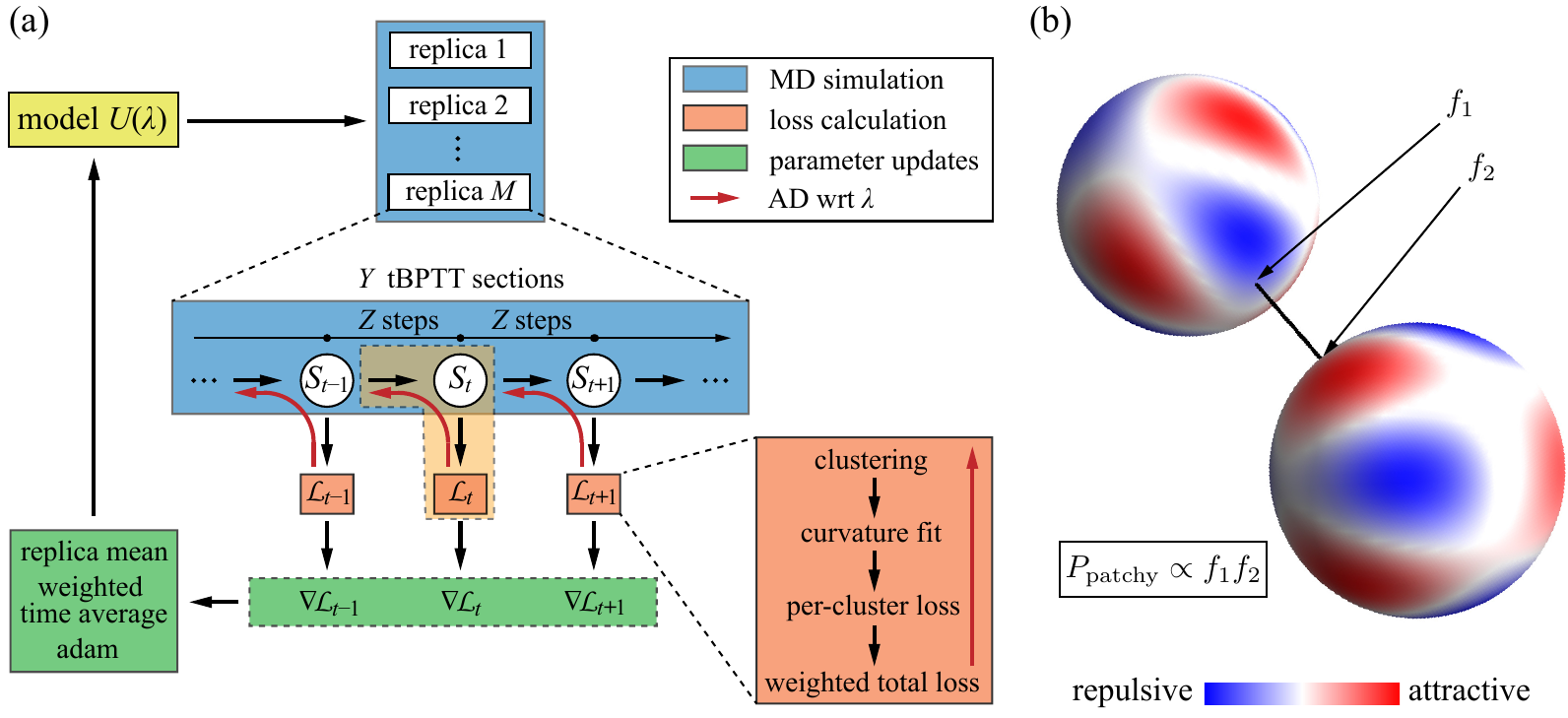}
\caption{a) Simulation scheme. MD simulations are preformed given an interaction model $U(\lambda)$ and multiple simulation replicas are run in parallel from different initial conditions to gather statistically significant trajectory sample. Each MD simulation consists of $Y$ smaller sections. Loss function and its gradient are calculated for each section separately. Contributions from all simulation replicas and simulation sections are then averaged to calculate the updates to interaction parameters $\lambda$. (b) Sketch of the patchy interaction model. Patches are described by spherical harmonic expansion coefficients and interaction between two particles depends only on patch values in the direction of the connecting line between particle centers. Two red patches with negative patch values attract each other, blue patches with positive values are repulsive, and there is no interaction between red and blue patches.}
\label{fig:simulation-scheme}
\end{figure*}

The schematics of our optimization algorithm are shown in Fig.~\ref{fig:simulation-scheme}. We consider an interaction model $U(\lambda)$ where the pair energy depends on a set of parameters $\lambda$. Given this model, we perform MD simulations of $M$ system replicas from different initial conditions to sample a sufficiently large portion of the configuration space, ensuring that the optimized potential performs well regardless of the starting configuration. MD simulations of self-assembly behavior typically require a high number of time steps to form large clusters which commonly leads to exploding gradients in AD algorithms when differentiating through long MD trajectory \cite{Metz2022}. We solve this by implementing a version of truncated backpropagation through time (tBPTT), originally developed for training of recurrent neural networks \cite{Sutskever2013}. Instead of traversing the entire MD simulation in a single backwards AD pass, the trajectory is broken down into $Y$ sections of $Z$ steps. The loss function $\mathcal{L}_t(\lambda)$ is calculated from the particle configuration $S_t$ at the end of each tBPTT section and its gradient over model parameters $\lambda$ is propagated only within that section. This results in $Y$ gradient estimates $\nabla\mathcal{L}_t$, capturing information from different time points in the self-assembly process while preventing numerical instabilities.

To support tBPTT, the loss function is constructed in a way that can be evaluated also for partially clustered states, thus conveying information about the assembly process. Particles are first grouped into clusters \cite{deOliveira2020}, then the Gauss-Newton minimization method is used to find a best fitting spherical surface to each cluster (no matter its shape, see SI for more details) and the resulting fit radii $R_k$ and fit residuals $\bm q_k$ are used to calculate the loss function,
\begin{equation}\label{eq:loss}
    \mathcal{L} = \frac{1}{N}\sum_{k=1}^{N'_\text{cl}} N_k \left( \left[\log\frac{R_k}{R_\text{target}}\right]^2 + \|\bm q_k \|_1 \right),
\end{equation}
where the sum runs over all clusters with more than three particles, $N_k>3$, and $\|\cdot\|_1$ is the Manhattan norm. Logarithmic map for the curvature deviation loss term is chosen to compensate for the fitted radii possibly being arbitrarily large if clusters are close to flat, and the residuals norm term penalizes structures that are not spherically shaped. Note that the loss function does not actively promote clustering so we must ensure that initial interaction parameters already result in self-assembled clusters. To determine the (approximate) loss gradient that defines the optimization step direction, a weighted average is taken over AD-calculated gradients $\nabla_\lambda\mathcal{L}_t$ from different BPTT sections, additionally averaged over all $M$ simulation replicas,
\begin{equation}\label{eq:loss_average}
  \langle \nabla_\lambda \mathcal{L} \rangle_t = \frac{1}{\sum_t h(t)} \sum_{t=0}^Y h(t) \nabla_\lambda\langle \mathcal{L}_t\rangle_M,
\end{equation}
where $h(t)$ is the time weight function. The averaged gradient value is then used as the input in the Adam optimizer \cite{Adam} with learning rate $\gamma$ to calculate parameter updates for the model $U(\lambda)$.


\subsection{Interaction model}

Patchy interactions can be described with a wide variety of models, ranging from the simple Kern-Frenkel model~\cite{Sciortino2009,Romano2010,Romano2011,Li2018,Chen2018} to models with bead-like interaction sites that allow for more intricate patch arrangements and can also describe an arbitrary number of different patch types \cite{Ma2021,Romano2020,Russo2022,King2024}. However, all of these models require some a priori knowledge on the number of patches/interaction beads which cannot be easily changed during AD-based optimization. We adopt an alternative approach proposed by \citet{Lieu2022} and describes patch distribution in terms of its spherical harmonic expansion coefficients,
\begin{equation}
    f(\theta,\varphi|\{c_{\ell m}\}) = \sum_{\ell=0}^{\ell_{max}} \sum_{m=-\ell}^\ell c_{\ell m} S_{\ell m}(\theta,\varphi),
\end{equation}
where $S_{\ell m}$ are real spherical harmonics. This allows for increased flexibility in patch arrangement during optimization, however, the interaction form taken by \citet{Lieu2022} does not account for interactions between components with different wave numbers $\ell$. We improve on this by generalizing the Kern-Frenkel interaction form and define interaction strength between two patchy particles $i$ and $j$ to be proportional to the product of patch values $f_i$ and $f_j$ in the direction of the connecting line between their centers, see Fig.~\ref{fig:simulation-scheme}b,
\begin{align}\label{eq:patchy-interaction}
    f_1 &= f(R^{-1}(q_i)\widehat{\bm{r}}_{ij}), \nonumber \\ 
    f_2 &= f(R^{-1}(q_j)\widehat{\bm{r}}_{ij}), \nonumber \\ 
    P_{\mathrm{patchy}}(\bm{r}_{ij},q_i,q_j|\{c_{\ell m}\}) &= -[\operatorname{sign}(f_1)+\operatorname{sign}(f_2)]f_1 f_2,
\end{align}
where $R(q_i)$ is the rotational matrix corresponding to the quaternion $q_i$ describing the orientation of particle $i$. The inclusion of the additional sign correcting factor allows for description of both repulsive and attractive patches with a single expansion. Patches of like signs attract for positive values and repel for negative values, while patches of the opposite signs do not interact. Note that the potential form without the sign correcting factor would resemble electrostatic interactions with attraction between patches of opposite sign~\cite{Noya2014}. 

The distance dependent part of the patchy interaction is described by the Morse potential,
\begin{equation}
    U_M(\bm{r}_{ij}| \epsilon_M, \alpha, \sigma_M) = \epsilon_M [1 - \exp(-\alpha (r_{ij} - \sigma_M))]^2 - \epsilon_M,
\end{equation}
that is commonly used to model attractive interactions between colloids~\cite{Oh2019,Goodrich2021,Lieu2022,King2024}. The hard core repulsion between particles is modeled by the Weeks-Chandler-Andersen (WCA) potential,
\begin{equation}
    U_{\mathrm{WCA}}(\bm r_{ij}|\epsilon,\sigma) = 
    \begin{cases} 
      4\epsilon \left[ \left( \frac{\sigma}{r_{ij}} \right)^{12} - \left( \frac{\sigma}{r_{ij}} \right)^6 \right] + \epsilon, & r_{ij} < r_c, \\
      0, & r \geq r_c,
    \end{cases}
\end{equation}
where $r_c = \sqrt[6]{2}\sigma$. For simplicity, we use $\sigma=\sigma_M=1$.

The main drawback of such orientational interaction construction is the pair energy degeneracy if any of the particles is rotated around the axis determined by the connecting line between particle centers. This significantly increases the number of competing structures that can arise during the assembly process, especially in 3D simulations. We observed no optimization towards consistently curved 2D structures in systems interacting only through the above patchy interaction model and using the loss defined in Eq.~\eqref{eq:loss}. Constraining the assembly with a loss function alone would require a large number of loss contributions and, in turn, complex multi-objective optimization procedures. An alternative approach is to promote 2D assembly by additional interaction contributions instead, e.g. it was shown that systems of planar quadrupole particles assemble into different 2D shapes \cite{VanWorkum2006}. Patchy interactions, added as a perturbative contribution, can then be used to control the curvature in such systems, however, we show in the SI that quadrupolar interactions preserve the energetic competition between aligned and anti-aligned particles, preventing the formation of structures with a consistent curvature.

We therefore construct a simplified potential that breaks the up-down particle symmetry, and can be considered as a model for additional more complex interparticle interactions, such as amphiphilic interactions or steric effects. The orientational part of the alignment interaction is inspired by the dipolar interaction form and defined as
\begin{equation}
    P_{\text{align}}(\hat{\bm{r}}_{ij}, \bm p_i, \bm p_j|\eta) = \bm p_i \cdot \bm p_j - \eta (\bm p_i \cdot \hat{\bm{r}}_{ij}) (\bm p_j \cdot \hat{\bm{r}}_{ij}),
\end{equation}
with a free parameter $\eta$. In the regime $\eta\in [1, 2]$, it acts as a stiffness parameter, as it can be linked to energy penalty for bending a layer of particles with parallel orientations. More complicated potentials with similar features are routinely used as coarse-grained models of cellular membranes \cite{Yuan2010,Jiang2022,Li2024} and comparable symmetry-breaking effects could be achieved by considering two patch types instead of one, however, this would add considerably to the simulation complexity. Note that the particle direction vector~$\bm p$ is aligned with the axis of the $S_{10}$ spherical harmonic for patch description. For simplicity, the distance dependence of the alignment interaction is taken the same as for the patchy interaction. Together with the hard core Weeks-Chandler-Andersen repulsion, the full interaction takes the form
\begin{equation}
    U = U_{WCA} + U_M (P_{\text{patchy}} + \mu P_{\text{align}}),
\end{equation}
where $\mu$ is a weight factor for the alignment interaction.

\subsection{Optimization example}

We demonstrate the effectiveness of the proposed optimization procedure by considering $N=100$ unit diameter particles ($\sigma=1$) in the simulation box at a density $\rho=0.03$ and target Gaussian curvature radius of $R_t = 3$, which is just slightly above the expected radius of a vesicle at this $N$ ($R_v\approx 2.8$). The training set consists of $M=16$ simulation replicas, run from random initial conditions for $Z\times Y=10^5$ time steps of size $\Delta t=10^{-4}$ where tBPTT truncation length is set to $Z=500$ steps, and we choose a linear time weight function, $h(t) = t$, to emphasize structures from later stages of the self-assembly process. The learning rate is set to $\gamma=0.01$. Unless stated otherwise, these parameter choices are used for all results shown in the paper. The simulation is written in JAX \cite{Jax} to make it end-to-end differentiable, with the MD part relying on the JAX-MD library \cite{jaxmd2020} with a custom implementation for anisotropic interaction. Additional simulation details are given in the SI.

The initial patch parameters $c_{\ell m}$, along with other non-optimized pair energy parameters, were selected to ensure that particles consistently formed clusters during the first MD iteration. Unless stated otherwise, all simulations were performed with the parameters $\epsilon=5.0$, $\epsilon_M=25.0$, $\alpha=1.0$, $\eta=1.6$, and $\mu=0.5$, with the energy unit defined such that $kT\equiv1$. To give the system minimal initial information and allow for unbiased exploration of the design space, the initial patch distribution was set to a quadrupolar configuration with only nonzero coefficient $c_{20}=1$, resulting in the formation of flat clusters during the first MD iteration.

\begin{figure*}[!ht]
\centering
\includegraphics[width=\textwidth]{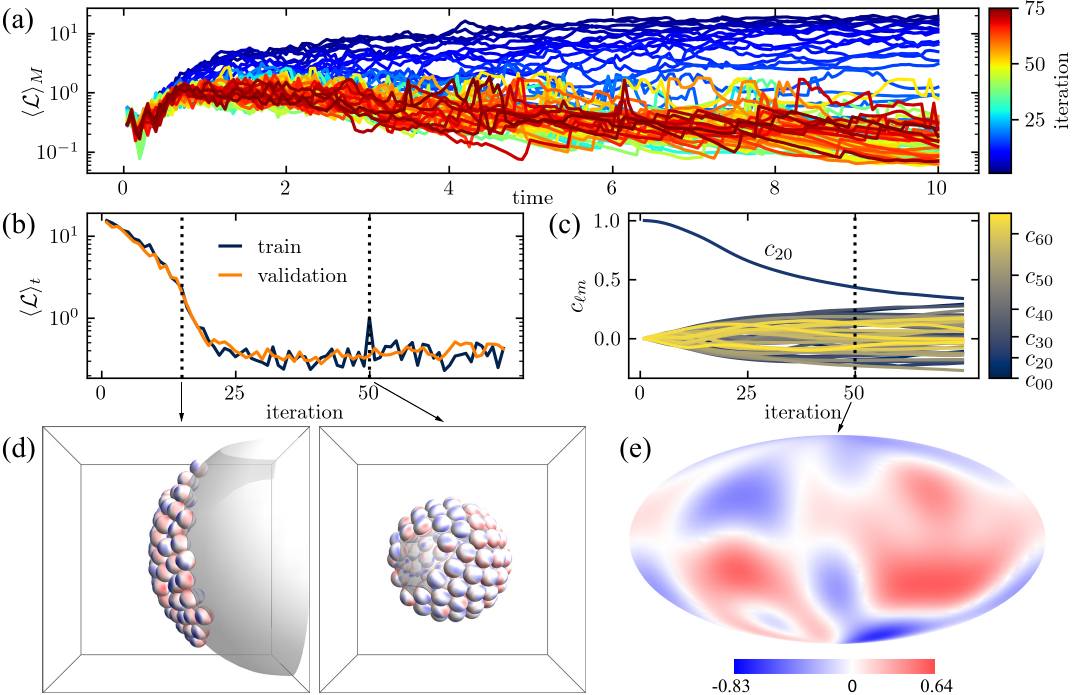}
\caption{Optimization demonstration for clusters with $R_t=3$. (a) Replica averaged loss values $\langle \mathcal{L}\rangle_M$ in dependence to MD time for all optimization iterations. Note that loss values are only calculated at the end of each tBPTT section and jumps are a consequence of merging smaller clusters. (b) Time averaged loss $\langle \mathcal{L}\rangle_t$ for both the training and validation initial configuration sets. (c) Change in patch coefficients $c_{\ell m}$ during optimization. (d) Example of cluster structure at final MD time with the fitted spherical surface for iterations 15 ($R=6.85$) and 50 ($R=2.90$), taken from the validation set of simulation replicas. (e) Patch distribution on a particle at iteration $i=50$, plotted in the Mollweide projection. The position of the north pole is defined by particle direction vector $\bm p$.}
\label{fig:opt-demo}
\end{figure*}

Figure \ref{fig:opt-demo}a shows the change in replica averaged loss curves, $\langle \mathcal{L}_t\rangle_M$, over optimization iterations. At the start of the assembly process, only a limited number of small clusters are already formed and the loss function is small by construction (low $N_k$ weights). During initial optimization iterations, $\mathcal{L}$ is increasing with time as interaction parameters support formation of flat clusters. A notable improvement in the final loss is observed already by iteration 25, and all later curves mostly vary around the same shape that shows a maximum around $t=1$, which is a consequence of thermal fluctuations significantly influencing the curvature of smaller clusters. Sudden jumps in the curves are caused by the merging of clusters which temporarily significantly increases fitted radius before the structure rearranges.

Further insight on the optimization performance is given by the time averaged loss $\langle \mathcal{L} \rangle_t$, defined equivalently to Eq.~\eqref{eq:loss_average}, i.e. we calculate a weighted average with respect to weights $h(t)$ of every curve in Fig.~\ref{fig:opt-demo}a and show the result in Fig.~\ref{fig:opt-demo}b. The improvement in $\langle \mathcal{L} \rangle_t$ during the first 30 iterations is approximately two orders of magnitude before a plateau is reached. We validate the optimization performance by performing $M=48$ additional self-assembly simulations from a separate set of initial conditions. The validation time-averaged loss $\langle \mathcal{L} \rangle_t$ closely follows the one for the training set, with a notable decrease in stochastic noise compared to training data and no observed overfitting, proving that the optimized patch patterns are not specific to training data. We demonstrate in Fig.~\ref{fig:opt-demo}d that optimization slowly decreases cluster curvature before a majority of clusters reach a curvature radius near the target $R_t=3$ around iteration $i=50$. The approach towards this value and the exact distribution of cluster radii after optimization is discussed further in Sec.~\ref{sec:target-radii}. Optimization beyond iteration $i=50$ does not yield any notable improvement in these distributions. 

The change in patch coefficients $\{c_{\ell m}\}$ during optimization is shown in Fig.~\ref{fig:opt-demo}c. The quick decrease in the initial quadrupolar $c_{20}$ contribution is partially aided by the coefficient rescaling factor that keeps average interaction magnitude constant (see SI) and a decrease in slope around iteration $i=25$ approximately coincides with the time averaged loss $\langle \mathcal{L} \rangle_t$ reaching a plateau. While coefficients change at a lower rate towards the end of the optimization, they do not reach convergence before we terminate the simulation, which can also be a consequence of the momentum term in the Adam optimizer \cite{Adam}. In Fig.~\ref{fig:opt-demo}e, we plot the patch distribution at iteration $i=50$, where cluster structure already matches the target to a great extent. The north-south symmetry (defined by particle direction $\bm p$) in patch distribution is broken, with stronger attractive regions in the south hemisphere. However, since the north-south symmetry is already broken by the alignment interactions, attractive patches in the northern hemisphere do not need to be completely absent to achieve consistent curved clusters. If no additional alignment interactions are present in the system, a more pronounced division between attractive and repulsive regions can be expected, as demonstrated in the SI.

\subsection{Hyperparameter analysis}\label{sec:hyper}

The success of the proposed tBPTT-based optimization scheme can depend to a large extent on the choice of hyperparameters, i.e. BPTT truncation length $Z$, time weight function $h(t)$, learning rate $\gamma$, and the number of simulation replicas $M$. We systematically explore the effects of each hyperparameter on optimization performance. Because performing a thorough grid search on the hyperparameter space is computationally not feasible, we vary each parameter separately while keeping the others at their baseline values.
\begin{figure*}[!ht]
\centering  
\includegraphics[width=\textwidth]{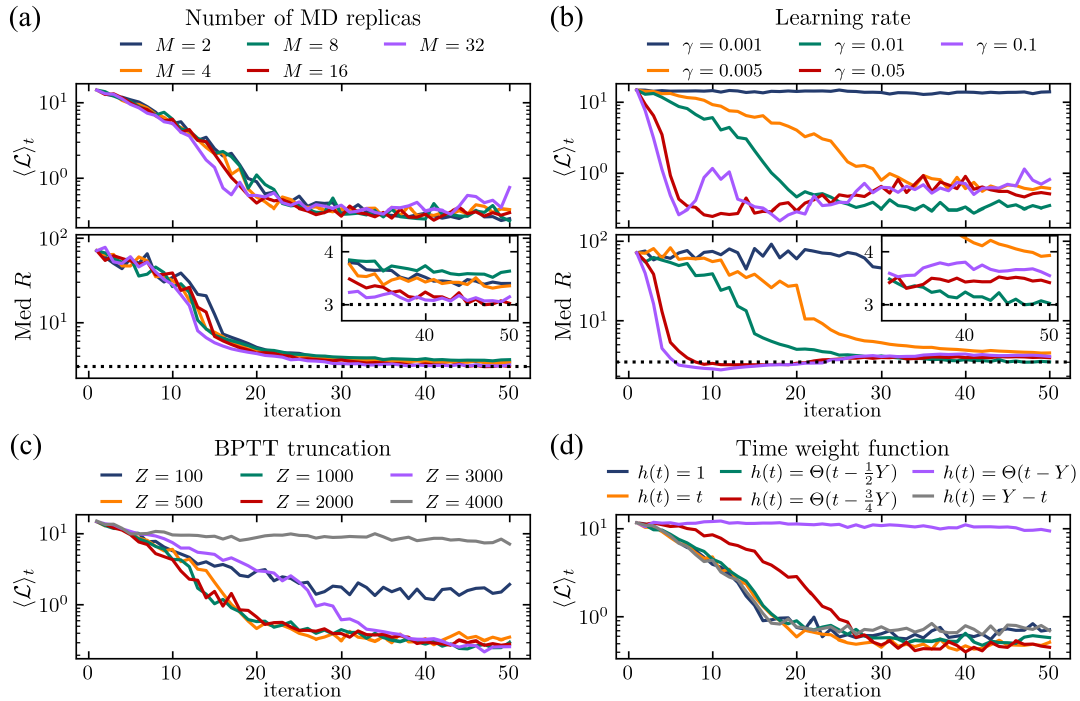}
\caption{Optimization curves for $\langle \mathcal{L}_t\rangle_M$ for changes in different hyperparameters: (a) number of MD simulation replicas, (b) learning rate for the Adam optimizer, (c) section length for BPTT truncation, and (d) time weight function for gradient averaging. To ensure that time weight curves are comparable in the plot even though the weights are changing, all curves in panel (d) were calculated using $h(t)=1$. Panels (a) and (b) additionally include plots that show the change in median cluster size $\mathrm{Med} R$, taking into account only clusters with $N_\mathrm{Cl}\geq 20$. The dotted line shows the target radius value $R_t=3$. The insets show the same $\mathrm{Med} R$ results during the final iterations on the linear scale.}
\label{fig:hyperparam}
\end{figure*}

The results are presented in Fig.~\ref{fig:hyperparam}. We first consider optimization at different numbers of simulation replicas $M$ where performance is similar across all tested values, i.e. validation $\langle \mathcal{L} \rangle_t$ curves coincide up to stochastic noise, including for the case of only two training simulation replicas. We confirm that a small number of replicas is sufficient to obtain generalizable optimized potentials by tracking the median cluster sizes at the final MD time (bottom panel of Fig.~\ref{fig:hyperparam}a). It is possible that the chaotic nature of the self-assembly process, where small changes in interaction parameters can significantly affect the MD trajectories, can offset the need for a large number of simulation replicas, at least for our optimization objective where all particles are interchangeable in the final structure. While the largest $M$ results are the closest to target radius $R_t=3$, the improvements are not monotonous, indicating at least some dependence on a specific set of training initial conditions. More details on cluster size distributions at final optimization iteration ($i=50$) are available in the SI.

Analysis of different learning rate values $\gamma$ shows that there exist a range of values where learning is optimal (Fig.~\ref{fig:hyperparam}b). If the learning rate is too small, optimization fails as changes in parameters are too small to meaningfully influence structural changes compared to other factors such as thermal fluctuations. Conversely, large learning rate values lead to very fast initial improvements, decreasing the cluster radius below $R_t$ at earlier iterations (bottom panel of Fig.~\ref{fig:hyperparam}b). In the following iterations, the radius gets over corrected and due to large parameter changes, the final results at iteration $i=50$ lie significantly above the target. Optimization examples with $\gamma = 0.05$ and $0.1$ work the best, however, results for $\gamma = 0.05$ would need more iterations to reach the target radius.

Similarly to learning rate, BPTT truncation length $Z$ also works only in a limited range, see Fig.~\ref{fig:hyperparam}c. At lower $Z$, the trajectory section over which we differentiate over is too short and the calculated gradient (i.e. its estimate compared to the true gradient over the entire trajectory) lacks important information on correlations over longer times. A working range of $Z$ is between $Z=500$ and $Z=2000$, where all simulations show comparable performance. The exact limit where exploding gradients start to interfere with optimization, lies around $Z=3000$ where initial optimization is slower, but by final iteration, the system still finds a solution. At $Z=4000$ and above, exploding gradients destroy the learning ability of the system.

We also consider different time weight functions $h(t)$ for gradient averaging to determine the importance of different stages in the self-assembly process for the optimization (Fig.~\ref{fig:hyperparam}d). Most of the trial function give very similar results, with the exception of the two step functions that only takes into account the final parts of the trajectory. Since most clusters are already formed at that point, we lose information on clustering dynamics and must instead rely mostly on thermal fluctuations for structural perturbations. Considering only the final tBPTT section does not provide sufficient data for optimization, and the last $Y/4$ sections show slower initial optimization when clusters are still close to flat. Conversely, weight functions that significantly consider starting tBPTT sections, i.e. $h(t)=1$ and $h(t)=Y-t$, also perform marginally worse compared to functions that elevate the second half of the tBPTT sections, demonstrating that the initial fast partial clustering is not representative of the final structures. There exist small differences in the distribution of cluster radii between different $h(t)$ that are further discussed in the SI.


\subsection{Different target radii}
\label{sec:target-radii}

\begin{figure}[!ht]
\centering  
\includegraphics[width=\columnwidth]{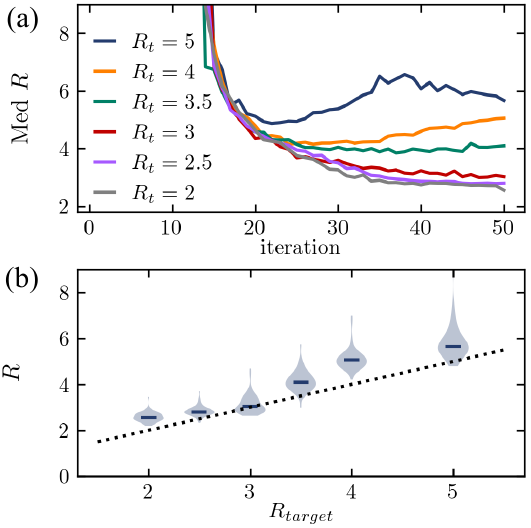}
\caption{Optimization performance with different target radii. (a) Median number of clusters at final MD time in dependence to optimization iteration. (b) Violin plot of distributions of cluster radii at iteration $i=50$. Blue lines indicate median cluster values. Only clusters with $N_\mathrm{cl} \geq 20$ are considered and for $R_t=4.$ and $R_t=5.$, we omit a small number ($\sim 5\%$) of clusters with $R>10$, resulting in more representative violin plots (see SI). Dotted line for $R=R_t$ indicates target results. Results for $R_t=2$ and $R_t=2.5$ were performed at decreased parameter values of $\epsilon=4$ and $\epsilon_M=20$.}
\label{fig:target-radius}
\end{figure}

We further investigate the performance of the proposed optimization scheme by considering different target radii of the self-assembled structures, with the results shown in Fig.~\ref{fig:target-radius}. During initial iterations, the structural improvements are independent of radius and individual curves only branch out when their median radii approach the respective target radius. Whereas the curves for lower target radii are monotonously descending (up to noise), we observe increasing behavior for higher target radii, away from the desired result. We hypothesize that because the rate of change in patch parameters (and consequently the median cluster radius) is still very large when these higher target values are approached, the system can over-correct for instances with smaller fitted radii, resulting in a bounce back. This could be mitigated by starting with a lower learning rate or implementing a decaying learning rate scheme in the simulation, however, this is beyond the scope of this paper.

Figure~\ref{fig:target-radius}b shows the distribution of cluster radii at the end of optimization. There is a general bias for cluster radii staying beyond the target radius, which is the consequence of the the described bounce back for larger target radii, as well as the construction of the loss function itself [Eq.~\eqref{eq:loss}] -- logarithmic map penalizes clusters with radii $R_t - \delta R$ more strictly than clusters with radii $R_t + \delta R$. Additionally, curvature radius is more sensitive to thermal fluctuations at smaller cluster radii where any positional displacements can significantly affect sphere fit.

The results for $R=3$ match the target value much better due to the tendency of the clusters with radii near the radius of a vesicle (with equal number of particles) to lower its energy by shrinking to close any existing holes. As stated above, for $N=100$ we get $R_v\approx 2.8$, which is also the approximate lower limit of the result distribution. This effect balances the tendency for larger radii that are preferred by the optimization algorithm alone. At target radii smaller than $R_v$, the same mechanism works in the other direction to again displace the distribution closer to $R_v$. Nevertheless, resulting distributions still monotonously follow any change in target radius. Consistently, self-limiting clusters are formed in the simulation box, which demonstrates that the optimization works also in larger systems with multiple final clusters.

\section{Discussion and conclusions}

We designed an AD-based optimization simulation to tackle general design problems in self-assembly processes. We demonstrated the effectiveness of the method by applying it to patch pattern optimization for formation of self-limiting clusters with a given target curvature radius in a one-component patchy particle system.

The biggest limitation of the proposed scheme lies in the loss function requirements. Besides its differentiability, relying on tBPTT calls for a loss function that can reasonably describe the progress of the assembly process at any MD time point, preventing us from considering only the final structure. Our optimization scheme also depends on numerous hyperparameters that significantly influence optimization performance and finding the optimal choices can be challenging. On the other hand, this leaves many opportunities for further optimization improvements. We here omitted fine-tuning of the Adam optimizer hyperparameters and did not investigate decreasing learning rate schedules which could be used to reach faster convergence as well as improved distributions of cluster radii to approach the target closer.

In this work, we only considered optimizing the patch parameters while keeping other interaction parameters constant. While this can be extended without adding any additional computational complexity, the MD interaction parameter values choices here were largely motivated by the need for a consistently fast clustering process. This also results in clusters where particles are not able to significantly redistribute. In longer simulations, we could work with slower processes at lower energies and optimize all interaction parameters in addition to patch coefficients to further improve the final cluster shapes.

Both the system size and the MD simulation length are limited by the relatively large computational cost inherent to differentiable MD simulations written in JAX, e.g. we omitted the use of neighbors list due to costly implementation. Still, further improvements in computational cost are possible. As the simulation is written purely in JAX, the entire code can be run on GPUs for a more parallel computation without any modification, which can significantly decrease simulation times with larger systems. Nevertheless, we opted for CPU only calculations, as multiple simulation replicas would require a large number of GPUs to surpass CPU performance in our demonstration case. 

The simulated system sizes mostly only allowed for one vesicle-like cluster to form in the simulation box, which partially limits the number of competing self-assembled structures that could arise in a larger system. This limits the transferability of the optimized patch distributions to larger systems, however, we demonstrated that the method also works when smaller clusters are targeted below the ``one cluster per simulation box'' limit. Nevertheless, while our example shows that a small number of training replicas is already sufficient for successful optimization, larger simulations with multiple self-assembled clusters in the simulation box or different target structures might still require a larger number of replicas to sufficiently sample all relevant self-assembly trajectories. We also acknowledge that the optimized patch patterns obtained in this work can be difficult to realize experimentally as they are inherently continuous, nevertheless, cutoff values can be chosen to discretize the patterns, making them more feasible to manufacture. The patchy interaction model used in the simulation could also be adapted to better match a specific experimental system.

The proposed algorithm significantly expands the possibilities for AD-based self-assembly inverse design. Future research could focus on expanding the optimization objectives to different curved surfaces, such as general quartic surfaces or even periodic structures, e.g. a gyroid. Designing these more complicated structures could require expanded interaction specificity, provided by more particle types or a larger number of distinct patch types. Such investigations could further advance our understanding of the fundamental principles behind the emerging curved geometry of assemblies and elucidate new ways of tuning properties of functional and biomimetic soft materials.

\section{Acknowledgments}
We acknowledge the support from Slovenian Research Agency (ARIS) under contracts no.\ P1-0099 (AG and SČ), J1-50006 (AG and SČ) and J1-3027 (SČ).

\bibliographystyle{apsrev4-2}
\bibliography{references}

\begin{thebibliography}{59}%
\makeatletter
\providecommand \@ifxundefined [1]{%
 \@ifx{#1\undefined}
}%
\providecommand \@ifnum [1]{%
 \ifnum #1\expandafter \@firstoftwo
 \else \expandafter \@secondoftwo
 \fi
}%
\providecommand \@ifx [1]{%
 \ifx #1\expandafter \@firstoftwo
 \else \expandafter \@secondoftwo
 \fi
}%
\providecommand \natexlab [1]{#1}%
\providecommand \enquote  [1]{``#1''}%
\providecommand \bibnamefont  [1]{#1}%
\providecommand \bibfnamefont [1]{#1}%
\providecommand \citenamefont [1]{#1}%
\providecommand \href@noop [0]{\@secondoftwo}%
\providecommand \href [0]{\begingroup \@sanitize@url \@href}%
\providecommand \@href[1]{\@@startlink{#1}\@@href}%
\providecommand \@@href[1]{\endgroup#1\@@endlink}%
\providecommand \@sanitize@url [0]{\catcode `\\12\catcode `\$12\catcode
  `\&12\catcode `\#12\catcode `\^12\catcode `\_12\catcode `\%12\relax}%
\providecommand \@@startlink[1]{}%
\providecommand \@@endlink[0]{}%
\providecommand \url  [0]{\begingroup\@sanitize@url \@url }%
\providecommand \@url [1]{\endgroup\@href {#1}{\urlprefix }}%
\providecommand \urlprefix  [0]{URL }%
\providecommand \Eprint [0]{\href }%
\providecommand \doibase [0]{https://doi.org/}%
\providecommand \selectlanguage [0]{\@gobble}%
\providecommand \bibinfo  [0]{\@secondoftwo}%
\providecommand \bibfield  [0]{\@secondoftwo}%
\providecommand \translation [1]{[#1]}%
\providecommand \BibitemOpen [0]{}%
\providecommand \bibitemStop [0]{}%
\providecommand \bibitemNoStop [0]{.\EOS\space}%
\providecommand \EOS [0]{\spacefactor3000\relax}%
\providecommand \BibitemShut  [1]{\csname bibitem#1\endcsname}%
\let\auto@bib@innerbib\@empty
\bibitem [{\citenamefont {Philp}\ and\ \citenamefont
  {Stoddart}(1996)}]{Philp1996}%
  \BibitemOpen
  \bibfield  {author} {\bibinfo {author} {\bibfnamefont {D.}~\bibnamefont
  {Philp}}\ and\ \bibinfo {author} {\bibfnamefont {J.~F.}\ \bibnamefont
  {Stoddart}},\ }\href {https://doi.org/10.1002/anie.199611541} {\bibfield
  {journal} {\bibinfo  {journal} {Angewandte Chemie International Edition in
  English}\ }\textbf {\bibinfo {volume} {35}},\ \bibinfo {pages} {1154–1196}
  (\bibinfo {year} {1996})}\BibitemShut {NoStop}%
\bibitem [{\citenamefont {Glotzer}\ and\ \citenamefont
  {Solomon}(2007)}]{Glotzer2007}%
  \BibitemOpen
  \bibfield  {author} {\bibinfo {author} {\bibfnamefont {S.~C.}\ \bibnamefont
  {Glotzer}}\ and\ \bibinfo {author} {\bibfnamefont {M.~J.}\ \bibnamefont
  {Solomon}},\ }\href {http://dx.doi.org/10.1038/nmat1949} {\bibfield
  {journal} {\bibinfo  {journal} {Nat. Mater.}\ }\textbf {\bibinfo {volume}
  {6}},\ \bibinfo {pages} {557} (\bibinfo {year} {2007})}\BibitemShut {NoStop}%
\bibitem [{\citenamefont {Étienne Duguet}\ \emph {et~al.}(2016)\citenamefont
  {Étienne Duguet}, \citenamefont {Hubert}, \citenamefont {Chomette},
  \citenamefont {Perro},\ and\ \citenamefont {Ravaine}}]{Duguet2016}%
  \BibitemOpen
  \bibfield  {author} {\bibinfo {author} {\bibnamefont {Étienne Duguet}},
  \bibinfo {author} {\bibfnamefont {C.}~\bibnamefont {Hubert}}, \bibinfo
  {author} {\bibfnamefont {C.}~\bibnamefont {Chomette}}, \bibinfo {author}
  {\bibfnamefont {A.}~\bibnamefont {Perro}},\ and\ \bibinfo {author}
  {\bibfnamefont {S.}~\bibnamefont {Ravaine}},\ }\href
  {http://dx.doi.org/10.1016/j.crci.2015.11.013} {\bibfield  {journal}
  {\bibinfo  {journal} {C. R. Chim.}\ }\textbf {\bibinfo {volume} {19}},\
  \bibinfo {pages} {173} (\bibinfo {year} {2016})}\BibitemShut {NoStop}%
\bibitem [{\citenamefont {Gong}\ \emph {et~al.}(2017)\citenamefont {Gong},
  \citenamefont {Hueckel}, \citenamefont {Yi},\ and\ \citenamefont
  {Sacanna}}]{Gong2017}%
  \BibitemOpen
  \bibfield  {author} {\bibinfo {author} {\bibfnamefont {Z.}~\bibnamefont
  {Gong}}, \bibinfo {author} {\bibfnamefont {T.}~\bibnamefont {Hueckel}},
  \bibinfo {author} {\bibfnamefont {G.-R.}\ \bibnamefont {Yi}},\ and\ \bibinfo
  {author} {\bibfnamefont {S.}~\bibnamefont {Sacanna}},\ }\href
  {https://doi.org/10.1038/nature23901} {\bibfield  {journal} {\bibinfo
  {journal} {Nature}\ }\textbf {\bibinfo {volume} {550}},\ \bibinfo {pages}
  {234} (\bibinfo {year} {2017})}\BibitemShut {NoStop}%
\bibitem [{\citenamefont {Moud}(2022)}]{Moud2022}%
  \BibitemOpen
  \bibfield  {author} {\bibinfo {author} {\bibfnamefont {A.~A.}\ \bibnamefont
  {Moud}},\ }\href {http://dx.doi.org/10.1016/j.colcom.2022.100595} {\bibfield
  {journal} {\bibinfo  {journal} {Colloids Interface Sci. Commun.}\ }\textbf
  {\bibinfo {volume} {47}},\ \bibinfo {pages} {100595} (\bibinfo {year}
  {2022})}\BibitemShut {NoStop}%
\bibitem [{\citenamefont {Bianchi}\ \emph {et~al.}(2017)\citenamefont
  {Bianchi}, \citenamefont {van Oostrum}, \citenamefont {Likos},\ and\
  \citenamefont {Kahl}}]{Bianchi2017}%
  \BibitemOpen
  \bibfield  {author} {\bibinfo {author} {\bibfnamefont {E.}~\bibnamefont
  {Bianchi}}, \bibinfo {author} {\bibfnamefont {P.~D.~J.}\ \bibnamefont {van
  Oostrum}}, \bibinfo {author} {\bibfnamefont {C.~N.}\ \bibnamefont {Likos}},\
  and\ \bibinfo {author} {\bibfnamefont {G.}~\bibnamefont {Kahl}},\ }\href
  {https://doi.org/10.1016/j.cocis.2017.03.010} {\bibfield  {journal} {\bibinfo
   {journal} {Current Opinion in Colloid \& Interface Science}\ }\textbf
  {\bibinfo {volume} {30}},\ \bibinfo {pages} {8} (\bibinfo {year}
  {2017})}\BibitemShut {NoStop}%
\bibitem [{\citenamefont {Zhang}\ \emph {et~al.}(2021)\citenamefont {Zhang},
  \citenamefont {Lyu}, \citenamefont {Xu}, \citenamefont {Mu},\ and\
  \citenamefont {Wang}}]{Zhang2021}%
  \BibitemOpen
  \bibfield  {author} {\bibinfo {author} {\bibfnamefont {T.}~\bibnamefont
  {Zhang}}, \bibinfo {author} {\bibfnamefont {D.}~\bibnamefont {Lyu}}, \bibinfo
  {author} {\bibfnamefont {W.}~\bibnamefont {Xu}}, \bibinfo {author}
  {\bibfnamefont {Y.}~\bibnamefont {Mu}},\ and\ \bibinfo {author}
  {\bibfnamefont {Y.}~\bibnamefont {Wang}},\ }\bibfield  {journal} {\bibinfo
  {journal} {Front. Phys.}\ }\textbf {\bibinfo {volume} {9}},\ \href
  {https://doi.org/10.3389/fphy.2021.672375} {10.3389/fphy.2021.672375}
  (\bibinfo {year} {2021})\BibitemShut {NoStop}%
\bibitem [{\citenamefont {Walther}\ and\ \citenamefont
  {Müller}(2013)}]{Walther2013}%
  \BibitemOpen
  \bibfield  {author} {\bibinfo {author} {\bibfnamefont {A.}~\bibnamefont
  {Walther}}\ and\ \bibinfo {author} {\bibfnamefont {A.~H.~E.}\ \bibnamefont
  {Müller}},\ }\href {https://doi.org/10.1021/cr300089t} {\bibfield  {journal}
  {\bibinfo  {journal} {Chem. Rev.}\ }\textbf {\bibinfo {volume} {113}},\
  \bibinfo {pages} {5194} (\bibinfo {year} {2013})}\BibitemShut {NoStop}%
\bibitem [{\citenamefont {Oh}\ \emph {et~al.}(2019)\citenamefont {Oh},
  \citenamefont {Lee}, \citenamefont {Glotzer}, \citenamefont {Yi},\ and\
  \citenamefont {Pine}}]{Oh2019}%
  \BibitemOpen
  \bibfield  {author} {\bibinfo {author} {\bibfnamefont {J.~S.}\ \bibnamefont
  {Oh}}, \bibinfo {author} {\bibfnamefont {S.}~\bibnamefont {Lee}}, \bibinfo
  {author} {\bibfnamefont {S.~C.}\ \bibnamefont {Glotzer}}, \bibinfo {author}
  {\bibfnamefont {G.-R.}\ \bibnamefont {Yi}},\ and\ \bibinfo {author}
  {\bibfnamefont {D.~J.}\ \bibnamefont {Pine}},\ }\href
  {http://dx.doi.org/10.1038/s41467-019-11915-1} {\bibfield  {journal}
  {\bibinfo  {journal} {Nat. Commun.}\ }\textbf {\bibinfo {volume} {10}},\
  \bibinfo {pages} {551} (\bibinfo {year} {2019})}\BibitemShut {NoStop}%
\bibitem [{\citenamefont {Noya}\ \emph {et~al.}(2014)\citenamefont {Noya},
  \citenamefont {Kolovos}, \citenamefont {Doppelbauer}, \citenamefont {Kahl},\
  and\ \citenamefont {Bianchi}}]{Noya2014}%
  \BibitemOpen
  \bibfield  {author} {\bibinfo {author} {\bibfnamefont {E.~G.}\ \bibnamefont
  {Noya}}, \bibinfo {author} {\bibfnamefont {I.}~\bibnamefont {Kolovos}},
  \bibinfo {author} {\bibfnamefont {G.}~\bibnamefont {Doppelbauer}}, \bibinfo
  {author} {\bibfnamefont {G.}~\bibnamefont {Kahl}},\ and\ \bibinfo {author}
  {\bibfnamefont {E.}~\bibnamefont {Bianchi}},\ }\href
  {http://dx.doi.org/10.1039/c4sm01559b} {\bibfield  {journal} {\bibinfo
  {journal} {Soft Matter}\ }\textbf {\bibinfo {volume} {10}},\ \bibinfo {pages}
  {8464} (\bibinfo {year} {2014})}\BibitemShut {NoStop}%
\bibitem [{\citenamefont {Morphew}\ \emph {et~al.}(2018)\citenamefont
  {Morphew}, \citenamefont {Shaw}, \citenamefont {Avins},\ and\ \citenamefont
  {Chakrabarti}}]{Morphew2018}%
  \BibitemOpen
  \bibfield  {author} {\bibinfo {author} {\bibfnamefont {D.}~\bibnamefont
  {Morphew}}, \bibinfo {author} {\bibfnamefont {J.}~\bibnamefont {Shaw}},
  \bibinfo {author} {\bibfnamefont {C.}~\bibnamefont {Avins}},\ and\ \bibinfo
  {author} {\bibfnamefont {D.}~\bibnamefont {Chakrabarti}},\ }\href
  {http://dx.doi.org/10.1021/acsnano.7b07633} {\bibfield  {journal} {\bibinfo
  {journal} {ACS Nano}\ }\textbf {\bibinfo {volume} {12}},\ \bibinfo {pages}
  {2355} (\bibinfo {year} {2018})}\BibitemShut {NoStop}%
\bibitem [{\citenamefont {Eslami}\ and\ \citenamefont
  {Müller-Plathe}(2024)}]{Eslami2024}%
  \BibitemOpen
  \bibfield  {author} {\bibinfo {author} {\bibfnamefont {H.}~\bibnamefont
  {Eslami}}\ and\ \bibinfo {author} {\bibfnamefont {F.}~\bibnamefont
  {Müller-Plathe}},\ }\href {https://doi.org/10.1002/smll.202306337}
  {\bibfield  {journal} {\bibinfo  {journal} {Small}\ }\textbf {\bibinfo
  {volume} {20}},\ \bibinfo {pages} {2306337} (\bibinfo {year}
  {2024})}\BibitemShut {NoStop}%
\bibitem [{\citenamefont {Zhang}\ and\ \citenamefont
  {Glotzer}(2004)}]{Zhang2004}%
  \BibitemOpen
  \bibfield  {author} {\bibinfo {author} {\bibfnamefont {Z.}~\bibnamefont
  {Zhang}}\ and\ \bibinfo {author} {\bibfnamefont {S.~C.}\ \bibnamefont
  {Glotzer}},\ }\href {http://dx.doi.org/10.1021/nl0493500} {\bibfield
  {journal} {\bibinfo  {journal} {Nano Lett.}\ }\textbf {\bibinfo {volume}
  {4}},\ \bibinfo {pages} {1407} (\bibinfo {year} {2004})}\BibitemShut
  {NoStop}%
\bibitem [{\citenamefont {Sherman}\ \emph {et~al.}(2020)\citenamefont
  {Sherman}, \citenamefont {Howard}, \citenamefont {Lindquist}, \citenamefont
  {Jadrich},\ and\ \citenamefont {Truskett}}]{Sherman2020}%
  \BibitemOpen
  \bibfield  {author} {\bibinfo {author} {\bibfnamefont {Z.~M.}\ \bibnamefont
  {Sherman}}, \bibinfo {author} {\bibfnamefont {M.~P.}\ \bibnamefont {Howard}},
  \bibinfo {author} {\bibfnamefont {B.~A.}\ \bibnamefont {Lindquist}}, \bibinfo
  {author} {\bibfnamefont {R.~B.}\ \bibnamefont {Jadrich}},\ and\ \bibinfo
  {author} {\bibfnamefont {T.~M.}\ \bibnamefont {Truskett}},\ }\href
  {http://dx.doi.org/10.1063/1.5145177} {\bibfield  {journal} {\bibinfo
  {journal} {J. Chem. Phys.}\ }\textbf {\bibinfo {volume} {152}},\ \bibinfo
  {pages} {928} (\bibinfo {year} {2020})}\BibitemShut {NoStop}%
\bibitem [{\citenamefont {Dijkstra}\ and\ \citenamefont
  {Luijten}(2021)}]{Dijkstra2021}%
  \BibitemOpen
  \bibfield  {author} {\bibinfo {author} {\bibfnamefont {M.}~\bibnamefont
  {Dijkstra}}\ and\ \bibinfo {author} {\bibfnamefont {E.}~\bibnamefont
  {Luijten}},\ }\href {http://dx.doi.org/10.1038/s41563-021-01014-2} {\bibfield
   {journal} {\bibinfo  {journal} {Nat. Mater.}\ }\textbf {\bibinfo {volume}
  {20}},\ \bibinfo {pages} {762} (\bibinfo {year} {2021})}\BibitemShut
  {NoStop}%
\bibitem [{\citenamefont {Kumar}\ \emph {et~al.}(2019)\citenamefont {Kumar},
  \citenamefont {Coli}, \citenamefont {Dijkstra},\ and\ \citenamefont
  {Sastry}}]{Kumar2019}%
  \BibitemOpen
  \bibfield  {author} {\bibinfo {author} {\bibfnamefont {R.}~\bibnamefont
  {Kumar}}, \bibinfo {author} {\bibfnamefont {G.~M.}\ \bibnamefont {Coli}},
  \bibinfo {author} {\bibfnamefont {M.}~\bibnamefont {Dijkstra}},\ and\
  \bibinfo {author} {\bibfnamefont {S.}~\bibnamefont {Sastry}},\ }\href
  {http://dx.doi.org/10.1063/1.5111492} {\bibfield  {journal} {\bibinfo
  {journal} {J. Chem. Phys.}\ }\textbf {\bibinfo {volume} {151}},\ \bibinfo
  {pages} {228301} (\bibinfo {year} {2019})}\BibitemShut {NoStop}%
\bibitem [{\citenamefont {Goodrich}\ \emph {et~al.}(2021)\citenamefont
  {Goodrich}, \citenamefont {King}, \citenamefont {Schoenholz}, \citenamefont
  {Cubuk},\ and\ \citenamefont {Brenner}}]{Goodrich2021}%
  \BibitemOpen
  \bibfield  {author} {\bibinfo {author} {\bibfnamefont {C.~P.}\ \bibnamefont
  {Goodrich}}, \bibinfo {author} {\bibfnamefont {E.~M.}\ \bibnamefont {King}},
  \bibinfo {author} {\bibfnamefont {S.~S.}\ \bibnamefont {Schoenholz}},
  \bibinfo {author} {\bibfnamefont {E.~D.}\ \bibnamefont {Cubuk}},\ and\
  \bibinfo {author} {\bibfnamefont {M.~P.}\ \bibnamefont {Brenner}},\ }\href
  {http://dx.doi.org/10.1073/pnas.2024083118} {\bibfield  {journal} {\bibinfo
  {journal} {Proc. Natl. Acad. Sci.}\ }\textbf {\bibinfo {volume} {118}},\
  \bibinfo {pages} {5595} (\bibinfo {year} {2021})}\BibitemShut {NoStop}%
\bibitem [{\citenamefont {Long}\ and\ \citenamefont
  {Ferguson}(2018)}]{Long2018}%
  \BibitemOpen
  \bibfield  {author} {\bibinfo {author} {\bibfnamefont {A.~W.}\ \bibnamefont
  {Long}}\ and\ \bibinfo {author} {\bibfnamefont {A.~L.}\ \bibnamefont
  {Ferguson}},\ }\href {http://dx.doi.org/10.1039/c7me00077d} {\bibfield
  {journal} {\bibinfo  {journal} {Mol. Syst. Des. Eng.}\ }\textbf {\bibinfo
  {volume} {3}},\ \bibinfo {pages} {49} (\bibinfo {year} {2018})}\BibitemShut
  {NoStop}%
\bibitem [{\citenamefont {Romano}\ \emph {et~al.}(2020)\citenamefont {Romano},
  \citenamefont {Russo}, \citenamefont {Kroc},\ and\ \citenamefont
  {Šulc}}]{Romano2020}%
  \BibitemOpen
  \bibfield  {author} {\bibinfo {author} {\bibfnamefont {F.}~\bibnamefont
  {Romano}}, \bibinfo {author} {\bibfnamefont {J.}~\bibnamefont {Russo}},
  \bibinfo {author} {\bibfnamefont {L.}~\bibnamefont {Kroc}},\ and\ \bibinfo
  {author} {\bibfnamefont {P.}~\bibnamefont {Šulc}},\ }\href
  {https://doi.org/10.1103/PhysRevLett.125.118003} {\bibfield  {journal}
  {\bibinfo  {journal} {Phys. Rev. Lett.}\ }\textbf {\bibinfo {volume} {125}},\
  \bibinfo {pages} {118003} (\bibinfo {year} {2020})}\BibitemShut {NoStop}%
\bibitem [{\citenamefont {Lieu}\ and\ \citenamefont
  {Yoshinaga}(2022)}]{Lieu2022}%
  \BibitemOpen
  \bibfield  {author} {\bibinfo {author} {\bibfnamefont {U.~T.}\ \bibnamefont
  {Lieu}}\ and\ \bibinfo {author} {\bibfnamefont {N.}~\bibnamefont
  {Yoshinaga}},\ }\href {http://dx.doi.org/10.1063/5.0072234} {\bibfield
  {journal} {\bibinfo  {journal} {J. Chem. Phys.}\ }\textbf {\bibinfo {volume}
  {156}},\ \bibinfo {pages} {6964} (\bibinfo {year} {2022})}\BibitemShut
  {NoStop}%
\bibitem [{\citenamefont {Rivera-Rivera}\ \emph {et~al.}(2023)\citenamefont
  {Rivera-Rivera}, \citenamefont {Moore},\ and\ \citenamefont
  {Glotzer}}]{Rivera-Rivera2023}%
  \BibitemOpen
  \bibfield  {author} {\bibinfo {author} {\bibfnamefont {L.~Y.}\ \bibnamefont
  {Rivera-Rivera}}, \bibinfo {author} {\bibfnamefont {T.~C.}\ \bibnamefont
  {Moore}},\ and\ \bibinfo {author} {\bibfnamefont {S.~C.}\ \bibnamefont
  {Glotzer}},\ }\href {https://doi.org/10.1039/D2SM01593E} {\bibfield
  {journal} {\bibinfo  {journal} {Soft Matter}\ }\textbf {\bibinfo {volume}
  {19}},\ \bibinfo {pages} {2726} (\bibinfo {year} {2023})}\BibitemShut
  {NoStop}%
\bibitem [{\citenamefont {King}\ \emph {et~al.}(2024)\citenamefont {King},
  \citenamefont {Du}, \citenamefont {Zhu}, \citenamefont {Schoenholz},\ and\
  \citenamefont {Brenner}}]{King2024}%
  \BibitemOpen
  \bibfield  {author} {\bibinfo {author} {\bibfnamefont {E.~M.}\ \bibnamefont
  {King}}, \bibinfo {author} {\bibfnamefont {C.~X.}\ \bibnamefont {Du}},
  \bibinfo {author} {\bibfnamefont {Q.-Z.}\ \bibnamefont {Zhu}}, \bibinfo
  {author} {\bibfnamefont {S.~S.}\ \bibnamefont {Schoenholz}},\ and\ \bibinfo
  {author} {\bibfnamefont {M.~P.}\ \bibnamefont {Brenner}},\ }\href
  {https://doi.org/10.1073/pnas.2311891121} {\bibfield  {journal} {\bibinfo
  {journal} {Proc. Natl. Acad. Sci.}\ }\textbf {\bibinfo {volume} {121}},\
  \bibinfo {pages} {e2311891121} (\bibinfo {year} {2024})}\BibitemShut
  {NoStop}%
\bibitem [{\citenamefont {Minkov}\ \emph {et~al.}(2020)\citenamefont {Minkov},
  \citenamefont {Williamson}, \citenamefont {Andreani}, \citenamefont {Gerace},
  \citenamefont {Lou}, \citenamefont {Song}, \citenamefont {Hughes},\ and\
  \citenamefont {Fan}}]{Minkov2020}%
  \BibitemOpen
  \bibfield  {author} {\bibinfo {author} {\bibfnamefont {M.}~\bibnamefont
  {Minkov}}, \bibinfo {author} {\bibfnamefont {I.~A.~D.}\ \bibnamefont
  {Williamson}}, \bibinfo {author} {\bibfnamefont {L.~C.}\ \bibnamefont
  {Andreani}}, \bibinfo {author} {\bibfnamefont {D.}~\bibnamefont {Gerace}},
  \bibinfo {author} {\bibfnamefont {B.}~\bibnamefont {Lou}}, \bibinfo {author}
  {\bibfnamefont {A.~Y.}\ \bibnamefont {Song}}, \bibinfo {author}
  {\bibfnamefont {T.~W.}\ \bibnamefont {Hughes}},\ and\ \bibinfo {author}
  {\bibfnamefont {S.}~\bibnamefont {Fan}},\ }\href
  {https://doi.org/10.1021/acsphotonics.0c00327} {\bibfield  {journal}
  {\bibinfo  {journal} {{ACS} Photonics}\ }\textbf {\bibinfo {volume} {7}},\
  \bibinfo {pages} {1729} (\bibinfo {year} {2020})}\BibitemShut {NoStop}%
\bibitem [{\citenamefont {Thaler}\ and\ \citenamefont
  {Zavadlav}(2021)}]{Thaler2021}%
  \BibitemOpen
  \bibfield  {author} {\bibinfo {author} {\bibfnamefont {S.}~\bibnamefont
  {Thaler}}\ and\ \bibinfo {author} {\bibfnamefont {J.}~\bibnamefont
  {Zavadlav}},\ }\href {http://dx.doi.org/10.1038/s41467-021-27241-4}
  {\bibfield  {journal} {\bibinfo  {journal} {Nat. Commun.}\ }\textbf {\bibinfo
  {volume} {12}},\ \bibinfo {pages} {230902} (\bibinfo {year}
  {2021})}\BibitemShut {NoStop}%
\bibitem [{\citenamefont {Engel}\ \emph {et~al.}(2023)\citenamefont {Engel},
  \citenamefont {Smith},\ and\ \citenamefont {Brenner}}]{Engel2023}%
  \BibitemOpen
  \bibfield  {author} {\bibinfo {author} {\bibfnamefont {M.~C.}\ \bibnamefont
  {Engel}}, \bibinfo {author} {\bibfnamefont {J.~A.}\ \bibnamefont {Smith}},\
  and\ \bibinfo {author} {\bibfnamefont {M.~P.}\ \bibnamefont {Brenner}},\
  }\href {http://dx.doi.org/10.1103/PhysRevX.13.041032} {\bibfield  {journal}
  {\bibinfo  {journal} {Phys. Rev. X}\ }\textbf {\bibinfo {volume} {13}},\
  \bibinfo {pages} {041032} (\bibinfo {year} {2023})}\BibitemShut {NoStop}%
\bibitem [{\citenamefont {Curatolo}\ \emph {et~al.}(2023)\citenamefont
  {Curatolo}, \citenamefont {Kimchi}, \citenamefont {Goodrich}, \citenamefont
  {Krueger},\ and\ \citenamefont {Brenner}}]{Curatolo2023}%
  \BibitemOpen
  \bibfield  {author} {\bibinfo {author} {\bibfnamefont {A.~I.}\ \bibnamefont
  {Curatolo}}, \bibinfo {author} {\bibfnamefont {O.}~\bibnamefont {Kimchi}},
  \bibinfo {author} {\bibfnamefont {C.~P.}\ \bibnamefont {Goodrich}}, \bibinfo
  {author} {\bibfnamefont {R.~K.}\ \bibnamefont {Krueger}},\ and\ \bibinfo
  {author} {\bibfnamefont {M.~P.}\ \bibnamefont {Brenner}},\ }\href
  {http://dx.doi.org/10.1038/s41467-023-43168-4} {\bibfield  {journal}
  {\bibinfo  {journal} {Nat. Commun.}\ }\textbf {\bibinfo {volume} {14}},\
  \bibinfo {pages} {3} (\bibinfo {year} {2023})}\BibitemShut {NoStop}%
\bibitem [{\citenamefont {Miskin}\ \emph {et~al.}(2016)\citenamefont {Miskin},
  \citenamefont {Khaira}, \citenamefont {de~Pablo},\ and\ \citenamefont
  {Jaeger}}]{Miskin2016}%
  \BibitemOpen
  \bibfield  {author} {\bibinfo {author} {\bibfnamefont {M.~Z.}\ \bibnamefont
  {Miskin}}, \bibinfo {author} {\bibfnamefont {G.}~\bibnamefont {Khaira}},
  \bibinfo {author} {\bibfnamefont {J.~J.}\ \bibnamefont {de~Pablo}},\ and\
  \bibinfo {author} {\bibfnamefont {H.~M.}\ \bibnamefont {Jaeger}},\ }\href
  {http://dx.doi.org/10.1073/pnas.1509316112} {\bibfield  {journal} {\bibinfo
  {journal} {Proc. Natl. Acad. Sci.}\ }\textbf {\bibinfo {volume} {113}},\
  \bibinfo {pages} {34} (\bibinfo {year} {2016})}\BibitemShut {NoStop}%
\bibitem [{\citenamefont {Xue}\ \emph {et~al.}(2022)\citenamefont {Xue},
  \citenamefont {Jin}, \citenamefont {Xu}, \citenamefont {Bai}, \citenamefont
  {He}, \citenamefont {Zhang}, \citenamefont {Cheng}, \citenamefont {Ji},
  \citenamefont {Pang}, \citenamefont {Shen}, \citenamefont {Song},
  \citenamefont {Shuai},\ and\ \citenamefont {Zhang}}]{Xue2022}%
  \BibitemOpen
  \bibfield  {author} {\bibinfo {author} {\bibfnamefont {Z.}~\bibnamefont
  {Xue}}, \bibinfo {author} {\bibfnamefont {T.}~\bibnamefont {Jin}}, \bibinfo
  {author} {\bibfnamefont {S.}~\bibnamefont {Xu}}, \bibinfo {author}
  {\bibfnamefont {K.}~\bibnamefont {Bai}}, \bibinfo {author} {\bibfnamefont
  {Q.}~\bibnamefont {He}}, \bibinfo {author} {\bibfnamefont {F.}~\bibnamefont
  {Zhang}}, \bibinfo {author} {\bibfnamefont {X.}~\bibnamefont {Cheng}},
  \bibinfo {author} {\bibfnamefont {Z.}~\bibnamefont {Ji}}, \bibinfo {author}
  {\bibfnamefont {W.}~\bibnamefont {Pang}}, \bibinfo {author} {\bibfnamefont
  {Z.}~\bibnamefont {Shen}}, \bibinfo {author} {\bibfnamefont {H.}~\bibnamefont
  {Song}}, \bibinfo {author} {\bibfnamefont {Y.}~\bibnamefont {Shuai}},\ and\
  \bibinfo {author} {\bibfnamefont {Y.}~\bibnamefont {Zhang}},\ }\bibfield
  {journal} {\bibinfo  {journal} {Science Advances}\ }\textbf {\bibinfo
  {volume} {8}},\ \href {https://doi.org/10.1126/sciadv.abm6922}
  {10.1126/sciadv.abm6922} (\bibinfo {year} {2022})\BibitemShut {NoStop}%
\bibitem [{\citenamefont {Mim}\ and\ \citenamefont {Unger}(2012)}]{Mim2012}%
  \BibitemOpen
  \bibfield  {author} {\bibinfo {author} {\bibfnamefont {C.}~\bibnamefont
  {Mim}}\ and\ \bibinfo {author} {\bibfnamefont {V.~M.}\ \bibnamefont
  {Unger}},\ }\href {https://doi.org/10.1016/j.tibs.2012.09.001} {\bibfield
  {journal} {\bibinfo  {journal} {Trends in Biochemical Sciences}\ }\textbf
  {\bibinfo {volume} {37}},\ \bibinfo {pages} {526–533} (\bibinfo {year}
  {2012})}\BibitemShut {NoStop}%
\bibitem [{\citenamefont {Shi}\ \emph {et~al.}(2023)\citenamefont {Shi},
  \citenamefont {Cannon}, \citenamefont {Curtis}, \citenamefont {Edelmaier},
  \citenamefont {Gladfelter},\ and\ \citenamefont {Nazockdast}}]{Shi2023}%
  \BibitemOpen
  \bibfield  {author} {\bibinfo {author} {\bibfnamefont {W.}~\bibnamefont
  {Shi}}, \bibinfo {author} {\bibfnamefont {K.~S.}\ \bibnamefont {Cannon}},
  \bibinfo {author} {\bibfnamefont {B.~N.}\ \bibnamefont {Curtis}}, \bibinfo
  {author} {\bibfnamefont {C.}~\bibnamefont {Edelmaier}}, \bibinfo {author}
  {\bibfnamefont {A.~S.}\ \bibnamefont {Gladfelter}},\ and\ \bibinfo {author}
  {\bibfnamefont {E.}~\bibnamefont {Nazockdast}},\ }\bibfield  {journal}
  {\bibinfo  {journal} {Proceedings of the National Academy of Sciences}\
  }\textbf {\bibinfo {volume} {120}},\ \href
  {https://doi.org/10.1073/pnas.2208253120} {10.1073/pnas.2208253120} (\bibinfo
  {year} {2023})\BibitemShut {NoStop}%
\bibitem [{\citenamefont {Roux}\ \emph {et~al.}(2021)\citenamefont {Roux},
  \citenamefont {Tozzi}, \citenamefont {Walani}, \citenamefont {Quiroga},
  \citenamefont {Zalvidea}, \citenamefont {Trepat}, \citenamefont {Staykova},
  \citenamefont {Arroyo},\ and\ \citenamefont {Roca-Cusachs}}]{Roux2021}%
  \BibitemOpen
  \bibfield  {author} {\bibinfo {author} {\bibfnamefont {A.-L.~L.}\
  \bibnamefont {Roux}}, \bibinfo {author} {\bibfnamefont {C.}~\bibnamefont
  {Tozzi}}, \bibinfo {author} {\bibfnamefont {N.}~\bibnamefont {Walani}},
  \bibinfo {author} {\bibfnamefont {X.}~\bibnamefont {Quiroga}}, \bibinfo
  {author} {\bibfnamefont {D.}~\bibnamefont {Zalvidea}}, \bibinfo {author}
  {\bibfnamefont {X.}~\bibnamefont {Trepat}}, \bibinfo {author} {\bibfnamefont
  {M.}~\bibnamefont {Staykova}}, \bibinfo {author} {\bibfnamefont
  {M.}~\bibnamefont {Arroyo}},\ and\ \bibinfo {author} {\bibfnamefont
  {P.}~\bibnamefont {Roca-Cusachs}},\ }\href
  {http://dx.doi.org/10.1038/s41467-021-26591-3} {\bibfield  {journal}
  {\bibinfo  {journal} {Nat. Commun.}\ }\textbf {\bibinfo {volume} {12}},\
  \bibinfo {pages} {191} (\bibinfo {year} {2021})}\BibitemShut {NoStop}%
\bibitem [{\citenamefont {Naskalska}\ \emph {et~al.}(2021)\citenamefont
  {Naskalska}, \citenamefont {Borzęcka-Solarz}, \citenamefont {Różycki},
  \citenamefont {Stupka}, \citenamefont {Bochenek}, \citenamefont {Pyza},\ and\
  \citenamefont {Heddle}}]{Naskalska2021}%
  \BibitemOpen
  \bibfield  {author} {\bibinfo {author} {\bibfnamefont {A.}~\bibnamefont
  {Naskalska}}, \bibinfo {author} {\bibfnamefont {K.}~\bibnamefont
  {Borzęcka-Solarz}}, \bibinfo {author} {\bibfnamefont {J.}~\bibnamefont
  {Różycki}}, \bibinfo {author} {\bibfnamefont {I.}~\bibnamefont {Stupka}},
  \bibinfo {author} {\bibfnamefont {M.}~\bibnamefont {Bochenek}}, \bibinfo
  {author} {\bibfnamefont {E.}~\bibnamefont {Pyza}},\ and\ \bibinfo {author}
  {\bibfnamefont {J.~G.}\ \bibnamefont {Heddle}},\ }\href
  {http://dx.doi.org/10.1021/acs.biomac.1c00630} {\bibfield  {journal}
  {\bibinfo  {journal} {Biomacromolecules}\ }\textbf {\bibinfo {volume} {22}},\
  \bibinfo {pages} {4146} (\bibinfo {year} {2021})}\BibitemShut {NoStop}%
\bibitem [{\citenamefont {Olson}\ \emph {et~al.}(2007)\citenamefont {Olson},
  \citenamefont {Hu},\ and\ \citenamefont {Keinan}}]{Olson2007}%
  \BibitemOpen
  \bibfield  {author} {\bibinfo {author} {\bibfnamefont {A.~J.}\ \bibnamefont
  {Olson}}, \bibinfo {author} {\bibfnamefont {Y.~H.~E.}\ \bibnamefont {Hu}},\
  and\ \bibinfo {author} {\bibfnamefont {E.}~\bibnamefont {Keinan}},\ }\href
  {http://dx.doi.org/10.1073/pnas.0709489104} {\bibfield  {journal} {\bibinfo
  {journal} {Proc. Natl. Acad. Sci.}\ }\textbf {\bibinfo {volume} {104}},\
  \bibinfo {pages} {20731} (\bibinfo {year} {2007})}\BibitemShut {NoStop}%
\bibitem [{\citenamefont {Matsuura}(2018)}]{Matsuura2018}%
  \BibitemOpen
  \bibfield  {author} {\bibinfo {author} {\bibfnamefont {K.}~\bibnamefont
  {Matsuura}},\ }\href {http://dx.doi.org/10.1039/c8cc03844a} {\bibfield
  {journal} {\bibinfo  {journal} {Chem. Commun.}\ }\textbf {\bibinfo {volume}
  {54}},\ \bibinfo {pages} {8944} (\bibinfo {year} {2018})}\BibitemShut
  {NoStop}%
\bibitem [{\citenamefont {Laniado}\ \emph {et~al.}(2021)\citenamefont
  {Laniado}, \citenamefont {Cannon}, \citenamefont {Miller}, \citenamefont
  {Sawaya}, \citenamefont {McNamara},\ and\ \citenamefont
  {Yeates}}]{Laniado2021}%
  \BibitemOpen
  \bibfield  {author} {\bibinfo {author} {\bibfnamefont {J.}~\bibnamefont
  {Laniado}}, \bibinfo {author} {\bibfnamefont {K.~A.}\ \bibnamefont {Cannon}},
  \bibinfo {author} {\bibfnamefont {J.~E.}\ \bibnamefont {Miller}}, \bibinfo
  {author} {\bibfnamefont {M.~R.}\ \bibnamefont {Sawaya}}, \bibinfo {author}
  {\bibfnamefont {D.~E.}\ \bibnamefont {McNamara}},\ and\ \bibinfo {author}
  {\bibfnamefont {T.~O.}\ \bibnamefont {Yeates}},\ }\href
  {http://dx.doi.org/10.1021/acsnano.0c07167} {\bibfield  {journal} {\bibinfo
  {journal} {ACS Nano}\ }\textbf {\bibinfo {volume} {15}},\ \bibinfo {pages}
  {4277} (\bibinfo {year} {2021})}\BibitemShut {NoStop}%
\bibitem [{\citenamefont {Fletcher}\ \emph {et~al.}(2013)\citenamefont
  {Fletcher}, \citenamefont {Harniman}, \citenamefont {Barnes}, \citenamefont
  {Boyle}, \citenamefont {Collins}, \citenamefont {Mantell}, \citenamefont
  {Sharp}, \citenamefont {Antognozzi}, \citenamefont {Booth}, \citenamefont
  {Linden}, \citenamefont {Miles}, \citenamefont {Sessions}, \citenamefont
  {Verkade},\ and\ \citenamefont {Woolfson}}]{Fletcher2013}%
  \BibitemOpen
  \bibfield  {author} {\bibinfo {author} {\bibfnamefont {J.~M.}\ \bibnamefont
  {Fletcher}}, \bibinfo {author} {\bibfnamefont {R.~L.}\ \bibnamefont
  {Harniman}}, \bibinfo {author} {\bibfnamefont {F.~R.~H.}\ \bibnamefont
  {Barnes}}, \bibinfo {author} {\bibfnamefont {A.~L.}\ \bibnamefont {Boyle}},
  \bibinfo {author} {\bibfnamefont {A.}~\bibnamefont {Collins}}, \bibinfo
  {author} {\bibfnamefont {J.}~\bibnamefont {Mantell}}, \bibinfo {author}
  {\bibfnamefont {T.~H.}\ \bibnamefont {Sharp}}, \bibinfo {author}
  {\bibfnamefont {M.}~\bibnamefont {Antognozzi}}, \bibinfo {author}
  {\bibfnamefont {P.~J.}\ \bibnamefont {Booth}}, \bibinfo {author}
  {\bibfnamefont {N.}~\bibnamefont {Linden}}, \bibinfo {author} {\bibfnamefont
  {M.~J.}\ \bibnamefont {Miles}}, \bibinfo {author} {\bibfnamefont {R.~B.}\
  \bibnamefont {Sessions}}, \bibinfo {author} {\bibfnamefont {P.}~\bibnamefont
  {Verkade}},\ and\ \bibinfo {author} {\bibfnamefont {D.~N.}\ \bibnamefont
  {Woolfson}},\ }\href {http://dx.doi.org/10.1126/science.1233936} {\bibfield
  {journal} {\bibinfo  {journal} {Science}\ }\textbf {\bibinfo {volume}
  {340}},\ \bibinfo {pages} {595} (\bibinfo {year} {2013})}\BibitemShut
  {NoStop}%
\bibitem [{\citenamefont {Evers}\ \emph {et~al.}(2016)\citenamefont {Evers},
  \citenamefont {Luiken}, \citenamefont {Bolhuis},\ and\ \citenamefont
  {Kegel}}]{Evers2016}%
  \BibitemOpen
  \bibfield  {author} {\bibinfo {author} {\bibfnamefont {C.~H.~J.}\
  \bibnamefont {Evers}}, \bibinfo {author} {\bibfnamefont {J.~A.}\ \bibnamefont
  {Luiken}}, \bibinfo {author} {\bibfnamefont {P.~G.}\ \bibnamefont
  {Bolhuis}},\ and\ \bibinfo {author} {\bibfnamefont {W.~K.}\ \bibnamefont
  {Kegel}},\ }\href {https://doi.org/10.1038/nature17956} {\bibfield  {journal}
  {\bibinfo  {journal} {Nature}\ }\textbf {\bibinfo {volume} {534}},\ \bibinfo
  {pages} {364–368} (\bibinfo {year} {2016})}\BibitemShut {NoStop}%
\bibitem [{\citenamefont {Wang}\ \emph {et~al.}(2021)\citenamefont {Wang},
  \citenamefont {Pan}, \citenamefont {Zhu}, \citenamefont {Xu}, \citenamefont
  {Tian}, \citenamefont {Endo}, \citenamefont {Sugiyama}, \citenamefont
  {Yang},\ and\ \citenamefont {Qian}}]{Wang2021}%
  \BibitemOpen
  \bibfield  {author} {\bibinfo {author} {\bibfnamefont {B.}~\bibnamefont
  {Wang}}, \bibinfo {author} {\bibfnamefont {R.}~\bibnamefont {Pan}}, \bibinfo
  {author} {\bibfnamefont {W.}~\bibnamefont {Zhu}}, \bibinfo {author}
  {\bibfnamefont {Y.}~\bibnamefont {Xu}}, \bibinfo {author} {\bibfnamefont
  {Y.}~\bibnamefont {Tian}}, \bibinfo {author} {\bibfnamefont {M.}~\bibnamefont
  {Endo}}, \bibinfo {author} {\bibfnamefont {H.}~\bibnamefont {Sugiyama}},
  \bibinfo {author} {\bibfnamefont {Y.}~\bibnamefont {Yang}},\ and\ \bibinfo
  {author} {\bibfnamefont {X.}~\bibnamefont {Qian}},\ }\href
  {http://dx.doi.org/10.1039/d0sm01817a} {\bibfield  {journal} {\bibinfo
  {journal} {Soft Matter}\ }\textbf {\bibinfo {volume} {17}},\ \bibinfo {pages}
  {1184} (\bibinfo {year} {2021})}\BibitemShut {NoStop}%
\bibitem [{\citenamefont {Metz}\ \emph {et~al.}(2022)\citenamefont {Metz},
  \citenamefont {Freeman}, \citenamefont {Schoenholz},\ and\ \citenamefont
  {Kachman}}]{Metz2022}%
  \BibitemOpen
  \bibfield  {author} {\bibinfo {author} {\bibfnamefont {L.}~\bibnamefont
  {Metz}}, \bibinfo {author} {\bibfnamefont {C.~D.}\ \bibnamefont {Freeman}},
  \bibinfo {author} {\bibfnamefont {S.~S.}\ \bibnamefont {Schoenholz}},\ and\
  \bibinfo {author} {\bibfnamefont {T.}~\bibnamefont {Kachman}},\ }\href@noop
  {} {\bibinfo {title} {Gradients are not all you need}} (\bibinfo {year}
  {2022}),\ \Eprint {https://arxiv.org/abs/2111.05803} {arXiv:2111.05803
  [cs.LG]} \BibitemShut {NoStop}%
\bibitem [{\citenamefont {Jadrich}\ \emph {et~al.}(2017)\citenamefont
  {Jadrich}, \citenamefont {Lindquist},\ and\ \citenamefont
  {Truskett}}]{Jadrich2017}%
  \BibitemOpen
  \bibfield  {author} {\bibinfo {author} {\bibfnamefont {R.~B.}\ \bibnamefont
  {Jadrich}}, \bibinfo {author} {\bibfnamefont {B.~A.}\ \bibnamefont
  {Lindquist}},\ and\ \bibinfo {author} {\bibfnamefont {T.~M.}\ \bibnamefont
  {Truskett}},\ }\href {http://dx.doi.org/10.1063/1.4981796} {\bibfield
  {journal} {\bibinfo  {journal} {J. Chem. Phys.}\ }\textbf {\bibinfo {volume}
  {146}},\ \bibinfo {pages} {4100} (\bibinfo {year} {2017})}\BibitemShut
  {NoStop}%
\bibitem [{\citenamefont {Bowick}\ and\ \citenamefont
  {Giomi}(2009)}]{Bowick2009}%
  \BibitemOpen
  \bibfield  {author} {\bibinfo {author} {\bibfnamefont {M.~J.}\ \bibnamefont
  {Bowick}}\ and\ \bibinfo {author} {\bibfnamefont {L.}~\bibnamefont {Giomi}},\
  }\href {https://doi.org/10.1080/00018730903043166} {\bibfield  {journal}
  {\bibinfo  {journal} {Adv. Phys.}\ }\textbf {\bibinfo {volume} {58}},\
  \bibinfo {pages} {449} (\bibinfo {year} {2009})}\BibitemShut {NoStop}%
\bibitem [{\citenamefont {Hübl}\ and\ \citenamefont
  {Goodrich}(2024)}]{Hubl2024}%
  \BibitemOpen
  \bibfield  {author} {\bibinfo {author} {\bibfnamefont {M.~C.}\ \bibnamefont
  {Hübl}}\ and\ \bibinfo {author} {\bibfnamefont {C.~P.}\ \bibnamefont
  {Goodrich}},\ }\href {https://doi.org/10.48550/arXiv.2405.13567} {\bibinfo
  {title} {Accessing semi-addressable self assembly with efficient structure
  enumeration}} (\bibinfo {year} {2024}),\ \Eprint
  {https://arxiv.org/abs/2405.13567 [cond-mat]} {2405.13567 [cond-mat]}
  \BibitemShut {NoStop}%
\bibitem [{\citenamefont {Russo}\ \emph {et~al.}(2022)\citenamefont {Russo},
  \citenamefont {Romano}, \citenamefont {Kroc}, \citenamefont {Sciortino},
  \citenamefont {Rovigatti},\ and\ \citenamefont {Šulc}}]{Russo2022}%
  \BibitemOpen
  \bibfield  {author} {\bibinfo {author} {\bibfnamefont {J.}~\bibnamefont
  {Russo}}, \bibinfo {author} {\bibfnamefont {F.}~\bibnamefont {Romano}},
  \bibinfo {author} {\bibfnamefont {L.}~\bibnamefont {Kroc}}, \bibinfo {author}
  {\bibfnamefont {F.}~\bibnamefont {Sciortino}}, \bibinfo {author}
  {\bibfnamefont {L.}~\bibnamefont {Rovigatti}},\ and\ \bibinfo {author}
  {\bibfnamefont {P.}~\bibnamefont {Šulc}},\ }\href
  {https://doi.org/10.1088/1361-648X/ac5479} {\bibfield  {journal} {\bibinfo
  {journal} {J. Phys.: Condens. Matter}\ }\textbf {\bibinfo {volume} {34}},\
  \bibinfo {pages} {354002} (\bibinfo {year} {2022})}\BibitemShut {NoStop}%
\bibitem [{\citenamefont {Bohlin}\ \emph {et~al.}(2023)\citenamefont {Bohlin},
  \citenamefont {Turberfield}, \citenamefont {Louis},\ and\ \citenamefont
  {Šulc}}]{Bohlin2023}%
  \BibitemOpen
  \bibfield  {author} {\bibinfo {author} {\bibfnamefont {J.}~\bibnamefont
  {Bohlin}}, \bibinfo {author} {\bibfnamefont {A.~J.}\ \bibnamefont
  {Turberfield}}, \bibinfo {author} {\bibfnamefont {A.~A.}\ \bibnamefont
  {Louis}},\ and\ \bibinfo {author} {\bibfnamefont {P.}~\bibnamefont {Šulc}},\
  }\href {https://doi.org/10.1021/acsnano.2c09677} {\bibfield  {journal}
  {\bibinfo  {journal} {{ACS} Nano}\ }\textbf {\bibinfo {volume} {17}},\
  \bibinfo {pages} {5387} (\bibinfo {year} {2023})}\BibitemShut {NoStop}%
\bibitem [{\citenamefont {Sutskever}(2013)}]{Sutskever2013}%
  \BibitemOpen
  \bibfield  {author} {\bibinfo {author} {\bibfnamefont {I.}~\bibnamefont
  {Sutskever}},\ }\emph {\bibinfo {title} {Training Recurrent Neural
  Networks}},\ \href@noop {} {Ph.D. thesis},\ \bibinfo {address} {CAN}
  (\bibinfo {year} {2013}),\ \bibinfo {note} {aAINS22066}\BibitemShut {NoStop}%
\bibitem [{\citenamefont {de~Oliveira}\ \emph {et~al.}(2020)\citenamefont
  {de~Oliveira}, \citenamefont {Khani}, \citenamefont {Maia},\ and\
  \citenamefont {Tavares}}]{deOliveira2020}%
  \BibitemOpen
  \bibfield  {author} {\bibinfo {author} {\bibfnamefont {F.~C.}\ \bibnamefont
  {de~Oliveira}}, \bibinfo {author} {\bibfnamefont {S.}~\bibnamefont {Khani}},
  \bibinfo {author} {\bibfnamefont {J.~M.}\ \bibnamefont {Maia}},\ and\
  \bibinfo {author} {\bibfnamefont {F.~W.}\ \bibnamefont {Tavares}},\ }\href
  {https://doi.org/10.1080/08927022.2020.1839661} {\bibfield  {journal}
  {\bibinfo  {journal} {Mol. Simul.}\ }\textbf {\bibinfo {volume} {46}},\
  \bibinfo {pages} {1453–1466} (\bibinfo {year} {2020})}\BibitemShut
  {NoStop}%
\bibitem [{\citenamefont {Kingma}\ and\ \citenamefont {Ba}(2017)}]{Adam}%
  \BibitemOpen
  \bibfield  {author} {\bibinfo {author} {\bibfnamefont {D.~P.}\ \bibnamefont
  {Kingma}}\ and\ \bibinfo {author} {\bibfnamefont {J.}~\bibnamefont {Ba}},\
  }\href@noop {} {\bibinfo {title} {Adam: A method for stochastic
  optimization}} (\bibinfo {year} {2017}),\ \Eprint
  {https://arxiv.org/abs/1412.6980} {arXiv:1412.6980 [cs.LG]} \BibitemShut
  {NoStop}%
\bibitem [{\citenamefont {Sciortino}\ \emph {et~al.}(2009)\citenamefont
  {Sciortino}, \citenamefont {Giacometti},\ and\ \citenamefont
  {Pastore}}]{Sciortino2009}%
  \BibitemOpen
  \bibfield  {author} {\bibinfo {author} {\bibfnamefont {F.}~\bibnamefont
  {Sciortino}}, \bibinfo {author} {\bibfnamefont {A.}~\bibnamefont
  {Giacometti}},\ and\ \bibinfo {author} {\bibfnamefont {G.}~\bibnamefont
  {Pastore}},\ }\href {http://dx.doi.org/10.1103/PhysRevLett.103.237801}
  {\bibfield  {journal} {\bibinfo  {journal} {Phys. Rev. Lett.}\ }\textbf
  {\bibinfo {volume} {103}},\ \bibinfo {pages} {237801} (\bibinfo {year}
  {2009})}\BibitemShut {NoStop}%
\bibitem [{\citenamefont {Romano}\ \emph {et~al.}(2010)\citenamefont {Romano},
  \citenamefont {Sanz},\ and\ \citenamefont {Sciortino}}]{Romano2010}%
  \BibitemOpen
  \bibfield  {author} {\bibinfo {author} {\bibfnamefont {F.}~\bibnamefont
  {Romano}}, \bibinfo {author} {\bibfnamefont {E.}~\bibnamefont {Sanz}},\ and\
  \bibinfo {author} {\bibfnamefont {F.}~\bibnamefont {Sciortino}},\ }\href
  {http://dx.doi.org/10.1063/1.3393777} {\bibfield  {journal} {\bibinfo
  {journal} {J. Chem. Phys.}\ }\textbf {\bibinfo {volume} {132}},\ \bibinfo
  {pages} {161} (\bibinfo {year} {2010})}\BibitemShut {NoStop}%
\bibitem [{\citenamefont {Romano}\ and\ \citenamefont
  {Sciortino}(2011)}]{Romano2011}%
  \BibitemOpen
  \bibfield  {author} {\bibinfo {author} {\bibfnamefont {F.}~\bibnamefont
  {Romano}}\ and\ \bibinfo {author} {\bibfnamefont {F.}~\bibnamefont
  {Sciortino}},\ }\href {http://dx.doi.org/10.1039/c0sm01494j} {\bibfield
  {journal} {\bibinfo  {journal} {Soft Matter}\ }\textbf {\bibinfo {volume}
  {7}},\ \bibinfo {pages} {5799} (\bibinfo {year} {2011})}\BibitemShut
  {NoStop}%
\bibitem [{\citenamefont {Li}\ \emph {et~al.}(2018)\citenamefont {Li},
  \citenamefont {Zhu}, \citenamefont {Lu},\ and\ \citenamefont {Sun}}]{Li2018}%
  \BibitemOpen
  \bibfield  {author} {\bibinfo {author} {\bibfnamefont {Z.-W.}\ \bibnamefont
  {Li}}, \bibinfo {author} {\bibfnamefont {Y.-L.}\ \bibnamefont {Zhu}},
  \bibinfo {author} {\bibfnamefont {Z.-Y.}\ \bibnamefont {Lu}},\ and\ \bibinfo
  {author} {\bibfnamefont {Z.-Y.}\ \bibnamefont {Sun}},\ }\href
  {http://dx.doi.org/10.1039/c8sm01631c} {\bibfield  {journal} {\bibinfo
  {journal} {Soft Matter}\ }\textbf {\bibinfo {volume} {14}},\ \bibinfo {pages}
  {7625} (\bibinfo {year} {2018})}\BibitemShut {NoStop}%
\bibitem [{\citenamefont {Chen}\ \emph {et~al.}(2018)\citenamefont {Chen},
  \citenamefont {Zhang},\ and\ \citenamefont {Torquato}}]{Chen2018}%
  \BibitemOpen
  \bibfield  {author} {\bibinfo {author} {\bibfnamefont {D.}~\bibnamefont
  {Chen}}, \bibinfo {author} {\bibfnamefont {G.}~\bibnamefont {Zhang}},\ and\
  \bibinfo {author} {\bibfnamefont {S.}~\bibnamefont {Torquato}},\ }\href
  {http://dx.doi.org/10.1021/acs.jpcb.8b05627} {\bibfield  {journal} {\bibinfo
  {journal} {J. Phys. Chem. B}\ }\textbf {\bibinfo {volume} {122}},\ \bibinfo
  {pages} {8462} (\bibinfo {year} {2018})}\BibitemShut {NoStop}%
\bibitem [{\citenamefont {Ma}\ \emph {et~al.}(2021)\citenamefont {Ma},
  \citenamefont {Aulicino},\ and\ \citenamefont {Ferguson}}]{Ma2021}%
  \BibitemOpen
  \bibfield  {author} {\bibinfo {author} {\bibfnamefont {Y.}~\bibnamefont
  {Ma}}, \bibinfo {author} {\bibfnamefont {J.~C.}\ \bibnamefont {Aulicino}},\
  and\ \bibinfo {author} {\bibfnamefont {A.~L.}\ \bibnamefont {Ferguson}},\
  }\href {https://doi.org/10.1021/acs.jpcb.0c08723} {\bibfield  {journal}
  {\bibinfo  {journal} {J. Phys. Chem. B}\ }\textbf {\bibinfo {volume} {125}},\
  \bibinfo {pages} {2398} (\bibinfo {year} {2021})}\BibitemShut {NoStop}%
\bibitem [{\citenamefont {Van~Workum}\ and\ \citenamefont
  {Douglas}(2006)}]{VanWorkum2006}%
  \BibitemOpen
  \bibfield  {author} {\bibinfo {author} {\bibfnamefont {K.}~\bibnamefont
  {Van~Workum}}\ and\ \bibinfo {author} {\bibfnamefont {J.~F.}\ \bibnamefont
  {Douglas}},\ }\href {https://doi.org/10.1103/PhysRevE.73.031502} {\bibfield
  {journal} {\bibinfo  {journal} {Phys. Rev. E}\ }\textbf {\bibinfo {volume}
  {73}},\ \bibinfo {pages} {031502} (\bibinfo {year} {2006})}\BibitemShut
  {NoStop}%
\bibitem [{\citenamefont {Yuan}\ \emph {et~al.}(010 )\citenamefont {Yuan},
  \citenamefont {Huang}, \citenamefont {Li}, \citenamefont {Lykotrafitis},\
  and\ \citenamefont {Zhang}}]{Yuan2010}%
  \BibitemOpen
  \bibfield  {author} {\bibinfo {author} {\bibfnamefont {H.}~\bibnamefont
  {Yuan}}, \bibinfo {author} {\bibfnamefont {C.}~\bibnamefont {Huang}},
  \bibinfo {author} {\bibfnamefont {J.}~\bibnamefont {Li}}, \bibinfo {author}
  {\bibfnamefont {G.}~\bibnamefont {Lykotrafitis}},\ and\ \bibinfo {author}
  {\bibfnamefont {S.}~\bibnamefont {Zhang}},\ }\href
  {https://doi.org/10.1103/PhysRevE.82.011905} {\bibfield  {journal} {\bibinfo
  {journal} {Phys. Rev. E}\ }\textbf {\bibinfo {volume} {82}},\ \bibinfo
  {pages} {011905} (\bibinfo {year} {2010-})}\BibitemShut {NoStop}%
\bibitem [{\citenamefont {Jiang}\ \emph {et~al.}(2022)\citenamefont {Jiang},
  \citenamefont {Harker-Kirschneck}, \citenamefont {Vanhille-Campos},
  \citenamefont {Pfitzner}, \citenamefont {Lominadze}, \citenamefont {Roux},
  \citenamefont {Baum},\ and\ \citenamefont {Šarić}}]{Jiang2022}%
  \BibitemOpen
  \bibfield  {author} {\bibinfo {author} {\bibfnamefont {X.}~\bibnamefont
  {Jiang}}, \bibinfo {author} {\bibfnamefont {L.}~\bibnamefont
  {Harker-Kirschneck}}, \bibinfo {author} {\bibfnamefont {C.}~\bibnamefont
  {Vanhille-Campos}}, \bibinfo {author} {\bibfnamefont {A.-K.}\ \bibnamefont
  {Pfitzner}}, \bibinfo {author} {\bibfnamefont {E.}~\bibnamefont {Lominadze}},
  \bibinfo {author} {\bibfnamefont {A.}~\bibnamefont {Roux}}, \bibinfo {author}
  {\bibfnamefont {B.}~\bibnamefont {Baum}},\ and\ \bibinfo {author}
  {\bibfnamefont {A.}~\bibnamefont {Šarić}},\ }\href
  {https://doi.org/10.1371/journal.pcbi.1010586} {\bibfield  {journal}
  {\bibinfo  {journal} {{PLOS} Computational Biology}\ }\textbf {\bibinfo
  {volume} {18}},\ \bibinfo {pages} {e1010586} (\bibinfo {year}
  {2022})}\BibitemShut {NoStop}%
\bibitem [{\citenamefont {Li}\ \emph {et~al.}(2024)\citenamefont {Li},
  \citenamefont {Yang}, \citenamefont {Tian}, \citenamefont {Li}, \citenamefont
  {Wang}, \citenamefont {Shi}, \citenamefont {Cui},\ and\ \citenamefont
  {Shan}}]{Li2024}%
  \BibitemOpen
  \bibfield  {author} {\bibinfo {author} {\bibfnamefont {S.}~\bibnamefont
  {Li}}, \bibinfo {author} {\bibfnamefont {H.}~\bibnamefont {Yang}}, \bibinfo
  {author} {\bibfnamefont {F.}~\bibnamefont {Tian}}, \bibinfo {author}
  {\bibfnamefont {W.}~\bibnamefont {Li}}, \bibinfo {author} {\bibfnamefont
  {H.}~\bibnamefont {Wang}}, \bibinfo {author} {\bibfnamefont {X.}~\bibnamefont
  {Shi}}, \bibinfo {author} {\bibfnamefont {Z.}~\bibnamefont {Cui}},\ and\
  \bibinfo {author} {\bibfnamefont {Y.}~\bibnamefont {Shan}},\ }\href
  {https://doi.org/10.1021/acsnano.4c04212} {\bibfield  {journal} {\bibinfo
  {journal} {{ACS} Nano}\ }\textbf {\bibinfo {volume} {18}},\ \bibinfo {pages}
  {27891} (\bibinfo {year} {2024})}\BibitemShut {NoStop}%
\bibitem [{\citenamefont {Bradbury}\ \emph {et~al.}(2018)\citenamefont
  {Bradbury}, \citenamefont {Frostig}, \citenamefont {Hawkins}, \citenamefont
  {Johnson}, \citenamefont {Leary}, \citenamefont {Maclaurin}, \citenamefont
  {Necula}, \citenamefont {Paszke}, \citenamefont {Vander{P}las}, \citenamefont
  {Wanderman-{M}ilne},\ and\ \citenamefont {Zhang}}]{Jax}%
  \BibitemOpen
  \bibfield  {author} {\bibinfo {author} {\bibfnamefont {J.}~\bibnamefont
  {Bradbury}}, \bibinfo {author} {\bibfnamefont {R.}~\bibnamefont {Frostig}},
  \bibinfo {author} {\bibfnamefont {P.}~\bibnamefont {Hawkins}}, \bibinfo
  {author} {\bibfnamefont {M.~J.}\ \bibnamefont {Johnson}}, \bibinfo {author}
  {\bibfnamefont {C.}~\bibnamefont {Leary}}, \bibinfo {author} {\bibfnamefont
  {D.}~\bibnamefont {Maclaurin}}, \bibinfo {author} {\bibfnamefont
  {G.}~\bibnamefont {Necula}}, \bibinfo {author} {\bibfnamefont
  {A.}~\bibnamefont {Paszke}}, \bibinfo {author} {\bibfnamefont
  {J.}~\bibnamefont {Vander{P}las}}, \bibinfo {author} {\bibfnamefont
  {S.}~\bibnamefont {Wanderman-{M}ilne}},\ and\ \bibinfo {author}
  {\bibfnamefont {Q.}~\bibnamefont {Zhang}},\ }\href
  {http://github.com/google/jax} {\bibinfo {title} {{JAX}: composable
  transformations of {P}ython+{N}um{P}y programs}} (\bibinfo {year}
  {2018})\BibitemShut {NoStop}%
\bibitem [{\citenamefont {Schoenholz}\ and\ \citenamefont
  {Cubuk}(2020)}]{jaxmd2020}%
  \BibitemOpen
  \bibfield  {author} {\bibinfo {author} {\bibfnamefont {S.~S.}\ \bibnamefont
  {Schoenholz}}\ and\ \bibinfo {author} {\bibfnamefont {E.~D.}\ \bibnamefont
  {Cubuk}},\ }in\ \href
  {https://papers.nips.cc/paper/2020/file/83d3d4b6c9579515e1679aca8cbc8033-Paper.pdf}
  {\emph {\bibinfo {booktitle} {Advances in Neural Information Processing
  Systems}}},\ Vol.~\bibinfo {volume} {33}\ (\bibinfo  {publisher} {Curran
  Associates, Inc.},\ \bibinfo {year} {2020})\BibitemShut {NoStop}%
\end{thebibliography}%


\begin{thebibliography}{1}

\bibitem{jaxmd2020}
Samuel~S. Schoenholz and Ekin~D. Cubuk.
\newblock Jax m.d. a framework for differentiable physics.
\newblock In {\em Advances in Neural Information Processing Systems},
  volume~33. Curran Associates, Inc., 2020.

\bibitem{deOliveira2020}
Fellipe~C. de~Oliveira, Shaghayegh Khani, João~M. Maia, and Frederico~W.
  Tavares.
\newblock Modified clustering algorithm for molecular simulation.
\newblock {\em Mol. Simul.}, 46(18):1453–1466, November 2020.

\bibitem{VanWorkum2006}
Kevin Van~Workum and Jack~F. Douglas.
\newblock Symmetry, equivalence, and molecular self-assembly.
\newblock {\em Phys. Rev. E}, 73:031502, Mar 2006.

\end{thebibliography}

\end{document}


\title{SUPPORTING INFORMATION: \\ Controlling curvature of self-assembling surfaces via patchy particle design}

\author{Andraž Gnidovec}
\author{Simon Čopar}
\affiliation{Faculty  of  Mathematics  and  Physics,  University  of  Ljubljana,  Ljubljana,  Slovenia\relax}

\maketitle

\section{Simulation details}

\begin{figure*}[!ht]
\centering  
\includegraphics[width=\textwidth]{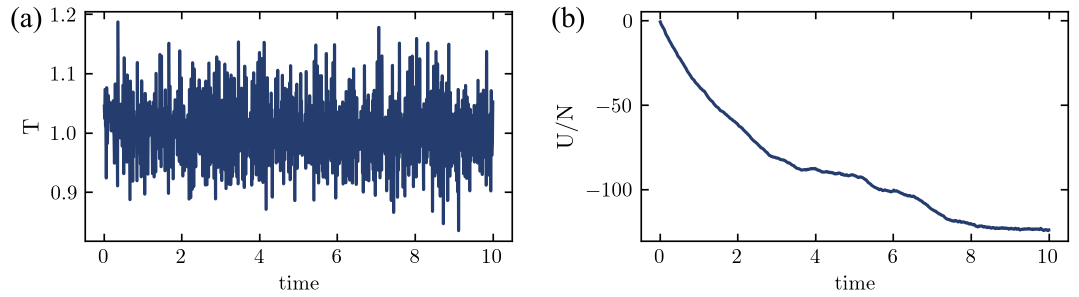}
\caption{MD simulation data from the optimization example shown in Fig.~2 in the main text. (a) Temperature and (b) potential energy per particle during MD simulation for a representative training set replica.}
\label{fig:md-log}
\end{figure*}

Molecular dynamics (MD) simulations were conducted using the JAX-MD package~\cite{jaxmd2020}, which inherently supports automatic differentiation. The simulations were performed with a fixed number of particles in a box with fixed dimensions and periodic boundary conditions. To control the temperature, we employ the Nos\'e-Hoover thermostat with a relaxation time of $\tau=50$ time steps, as it demonstrated better temperature stability compared to the Langevin thermostat. The MD simulations were initialized with non-overlapping particle configurations and random orientations, mimicking a rapid quench from a high-temperature state to a low-temperature state. The simulated self-assembly process is thus inherently non-equilibrium. Nevertheless, we show in Fig.~\ref{fig:md-log} that the temperature relaxes to the final temperature within the first $\sim 10^4$ time steps (ie. when $t\sim 1$ as $\Delta t=10^{-4}$) where gradient time weights $h(t)=t$ are small.

Because the defined loss function does not actively promote cluster formation, we instead rely on the initial attraction between particles in the model being approximately preserved during the optimization to keep the same clustering rate. To achieve this, coefficients $c_{\ell m}$ are rescaled at every iteration so that the product term $f_1 f_2$ in the patchy interaction definition [Eq.~(4) in the main text], averaged over all mutual orientations of two particles, remains constant. This condition reduces to demanding
\begin{equation}\label{eq:patch-renormalization}
\sum_{\ell m} |c_{\ell m}|^2 = 1.
\end{equation}
Without this rescaling, the optimization can favor faster clustering rates by increasing energy scales in the system, which decreases thermostat effectiveness.

If the initial interaction parameter values fail to support clustering, an additional loss term should be introduced to guide the system toward self-assembly. However, this loss cannot rely directly on the number of clusters, which as a discrete variable that cannot be differentiated, and should instead be constructed based on particle spatial distribution.

Even though truncated backpropagation through time (tBPTT) reduces RAM usage by performing reverse mode AD only on a smaller section of an MD trajectory, memory requirements still remain a limiting factor. A significant amount of intermediate calculations must be kept in memory due to a large number of particles and complexity in evaluating the patch values from the spherical harmonics expansion in the energy calculation. To mitigate this, gradient checkpointing is implemented at every energy evaluation, which represents a trade-off between memory requirements and additional computational cost. Different choices of checkpoints, eg. after a small number of time steps, showed slower performance in our testing.

\section{Cluster curvature determination via sphere fitting}

We determine the curvature of clusters by fitting a sphere to the point cloud data of each cluster which is more robust to positional imperfections and thermal fluctuations compared to local curvature estimation methods. We use the Gauss-Newton nonlinear least squares method to minimize a residual function $ ||\bm{q}_k||^2 $, where $\bm{q}_k$ measures linear distances between the fitted sphere and the positions of particles within each cluster $k$. The particle positions are contained in an $N_k \times 3$ matrix $\bm{Q}_k$, and the residuals are defined as
\begin{equation}
\bm{q}_k(R_\text{fit}, \bm{r}_0| \bm{Q}_k(\lambda)) = \sqrt{(\bm{Q}_k - \bm{1}_{N_k} \bm{r}_0)\cdot(\bm{Q}_k - \bm{1}_{N_k} \bm{r}_0)^T} - R_\text{fit} \bm{1}_{N_k},
\end{equation}
where $\bm{1}_{N_k}$ is a vector of ones with length $N_k$. The minimization is carried out with respect to the curvature radius $R_\text{fit}$ and the sphere center position $\bm{r}_0$.

Although the clustering algorithm \cite{deOliveira2020} considers periodic boundary conditions (PBCs) when grouping particles into clusters, the entire cluster must be mapped into the center of the simulation box before the sphere fitting. Advanced algorithms for reconstructing contiguous clusters are impractical to implement as a part of an AD-based optimization as they introduce additional complexity and time cost to the simulation. We therefore rely on a simple cluster reconstruction algorithm that rearranges the subclusters, identified without considering PBCs, around the common center of mass which is calculated by mapping the particle positions onto cyclic coordinates. We note that for highly elongated clusters, this algorithm may fail to produce a contiguous cluster without PBCs and dislocated cluster parts can skew the sphere fit, resulting in converged solutions that do not accurately reflect the curved surface but such errors are rare for results presented in the main text. Gradient averaging over time and over system replicas makes there contributions irrelevant.

The effectiveness of the fitting algorithm depends heavily on the initial guesses for $\bm{r}_0$ and, to a lesser extent, $R_\text{fit}$. If the initial sphere center position is placed on the ``outside'' of the optimal fitting sphere, the algorithm often fails to converge to the correct solution. To address this, an additional step is performed before sphere fitting where we find the best fitting plane for each cluster. The normal vector $\bm{n}$ of this plane is determined by the eigenvector corresponding to the smallest eigenvalue of the $3 \times 3$ matrix:
\begin{equation}
M = \frac{1}{N_k} (\bm{Q}_k - \bm{1}_{N_k} \bm{r}_{\text{CM},k})^T \cdot (\bm{Q}_k - \bm{1}_{N_k} \bm{r}_{\text{CM},k}),
\end{equation}
where $\bm{r}_{\text{CM},k}$ represents the center of mass of cluster $k$. Since the curvature direction is initially unknown, two sphere-fitting attempts are performed, each starting from an initial center point $\bm{r}_0$ on the opposite sides of the plane defined by $\bm{n}$. Typically, at least one of these attempts converges, and when both succeed, the solution with the lowest mean residual is selected. Nevertheless, in some cases, neither fitting attempt converges within the finite iteration limit, yielding excessively large curvature radius values. These cases are handled within the loss function calculation by imposing a cutoff at $R_\text{fit}=100R_t$. Flat clusters at the initial stages of optimization may also fall outside this cutoff, but such cases are rare and do not significantly impact the overall results.

\section{Closed clusters}

\begin{figure*}[!ht]
\centering  
\includegraphics[width=0.8\textwidth]{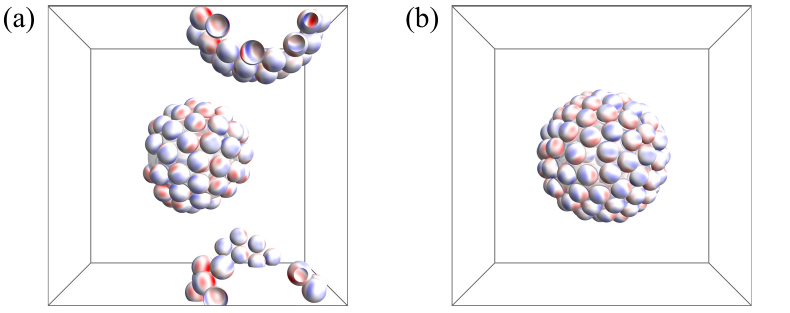}
\caption{Examples of vesicle formation for target radius (a) $R_t=2$, showing a vesicle with $N=59$ particles and curvature radius $R=2.28$, and (b) $R_t=3$, showing a vesicle with $N=100$ particles and $R=2.78$. In panel (a) the remaining $N=41$ particles form a vesicle with $R=2.54$ (non-contiguous due to periodic boundary conditions).} 
\label{fig:vesicles}
\end{figure*}

In Fig.~2 of the main paper, we show a representative cluster structure for $R_t=3$ that has the shape of an open vesicle. This vesicle cannot close in the simulations as the number of particles is kept constant and pair interactions are set to be strong enough to prevent collective structural rearrangements of particles in the vesicle. Some closed vesicles still form for certain MD replicas when the target radius is similar to the radius of a closed vesicle at a given number of particles in the simulation box, but usually still exhibit regions where particles are not fully close packed, as shown in Fig.~\ref{fig:vesicles}b. For target radii that support multiple vesicles in the simulation box, it is harder to obtain closed structures as the fast assembly process in our simulations quickly divides the particles between different proto-clusters that interact with each other very weakly. They thus merge only rarely and also cannot obtain new individual particles that would allow the vesicle to close, resulting in a number or partially formed vesicles (most commonly two at the end of our simulation time). Figure~\ref{fig:vesicles}a shows one of a few examples where one of the smaller clusters is almost closed.

\section{Distribution of radii}

\begin{figure*}[!ht]
\centering  
\includegraphics[width=0.8\textwidth]{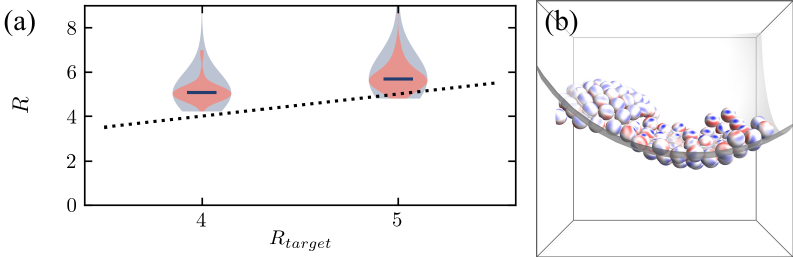}
\caption{(a) Comparison between violin plots for validation set curvature radius distributions considering all assembled clusters with $N_k\geq 20$ (blue) and excluding clusters with $R\geq 10$ (orange). (b) Example of a cluster with $R=$ during optimization with $R_\mathrm{target}=5$. }
\label{fig:rdist-remove-large}
\end{figure*}

In this section, we show additional results on the distributions of radii for the results shown in the main paper. Figure~\ref{fig:rdist-remove-large}a shows the change in violin plot representations from Fig. 4 in the main text if clusters with $R\geq 10$ are kept in the distribution or removed during the analysis. Such clusters are formed when two smaller clusters merge with opposite surface orientations (as defined by alignment interaction directions $\bm p_i$), resulting in a cluster that is not consistently curved. At larger target radii, such clusters often span over the entire simulation box in one dimension and their ends can become connected because of PBCs, which additionally stabilizes the structure and makes particles unable to reorganize into a shape of constant curvature (see Fig.~\ref{fig:rdist-remove-large}b). While these cases are rare (2-3 instances among 48 validation MD replicas), they significantly affect the violin plots and we deem plots without these results to be more representative of the distribution of cluster radii.

\begin{figure*}[!ht]
\centering  
\includegraphics[width=\textwidth]{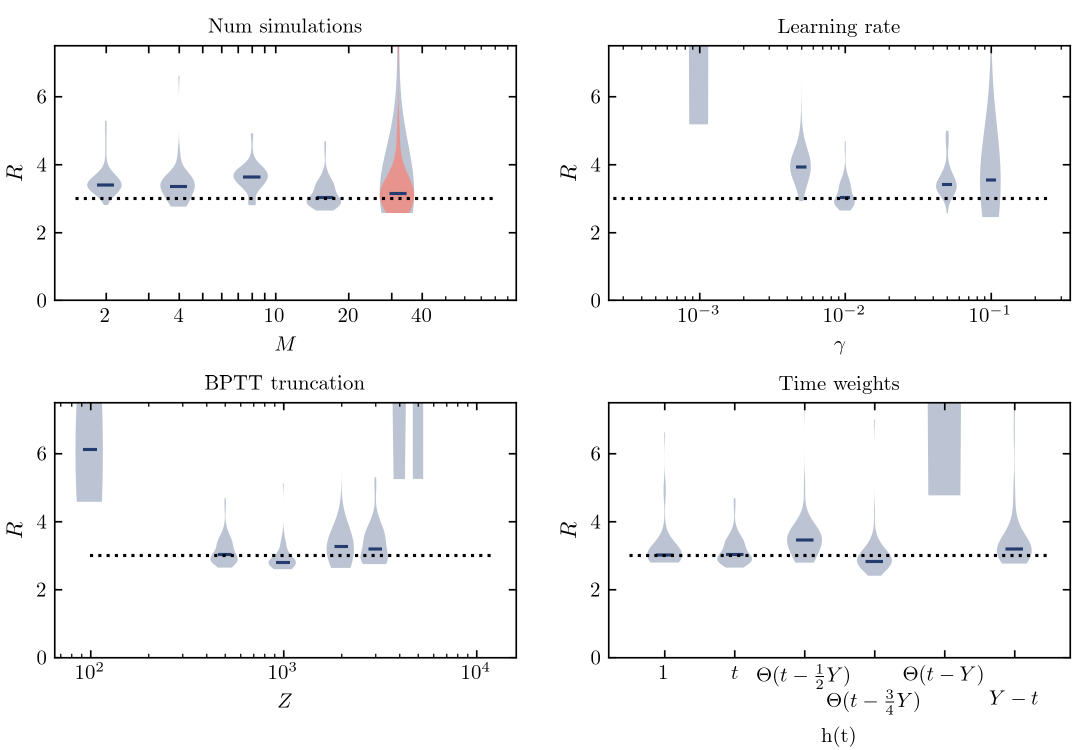}
\caption{Distribution of validation set cluster radii at iteration $i=50$ for optimization performed with different hyperparameter values. Blue lines inside the violin plots mark median radius values. In the panel comparing different numbers of MD simulation replicas, the orange violin plot only considers clusters with $R<10$. The dotted line indicates the target radius value $R_\mathrm{target}=3$.}
\label{fig:hyperparam-rdist}
\end{figure*}

We further show in Fig.~\ref{fig:hyperparam-rdist} the distribution of radii for the hyperparameter comparison results shown in Fig.~3 in the main text. Changing the number of training MD simulation replicas does not yield monotonous improvements and results at $M=8$ deviate from the target radius more than results for smaller $M$ which can be a consequence of specifics in the training set configurations. Distributions for different learning rate and BPTT truncation values mirror the findings presented in the main paper. Time weight function only has a small effect on final distribution of radii, with the exception of the $\Theta(t-Y)$ where optimization fails.

\section{Quadrupolar interactions}

Patchy interaction model as proposed in the main text lacks specificity to enable optimization for target curved structures on its own. The constraint to form 2D structures during the initial MD simulation, along with the pair energy degeneracy on single particle rotations, make it very hard for the learning algorithm to change the patch distribution to both preserve local 2D order as well as induce consistent cluster curvature. Considering only patchy interactions and starting from a belt-like patch distribution, we were not able to find a parameter regime where optimization leads to curved 2D sstructures. In the main text, we show that breaking the pair energy dipolar symmetry is enough to guarantee successful optimization. Here, we show that optimization is possible also in cases where some additional interaction promotes the formation of surface-like structures. Specifically, we consider quadrupolar interactions between planar quadrupoles that were shown to assemble into 2D structures \cite{VanWorkum2006}. Interaction between planar quadrupoles is given by
\begin{equation}
    U_\text{quad}(Q_i,Q_j,\bm{r}_{ij}) = Q_i:T(\bm{r}_{ij}):Q_j,
\end{equation}
where $:$ denotes double contraction and
\begin{equation}\label{eq:quad_int_tensor}
T_{ijkl}(\bm{r}) = \frac{1}{3r^5} \left[35\frac{x_i x_j x_k x_l}{r^4} - 20 \frac{x_j x_k \delta_{il}}{r^2} + 2 \delta_{ik}\delta_{jl}\right].
\end{equation}
In the particle frame, quadrupoles are placed onto the equatorial plane and in the simulation frame, quadrupole tensors depend on particle orientations, $Q_i=Q_i(q_i)$.The parametrization of quadrupole tensors in the rescaled form is \cite{VanWorkum2006}
\begin{equation}
    Q=\frac{3}{4} q_0 \sqrt{\epsilon \sigma^5} \operatorname{diag}(1, -1, 0),
\end{equation}
with a dimensionless quadrupolar strength $q_0$. 

\begin{figure*}[!ht]
\centering  
\includegraphics[width=\textwidth]{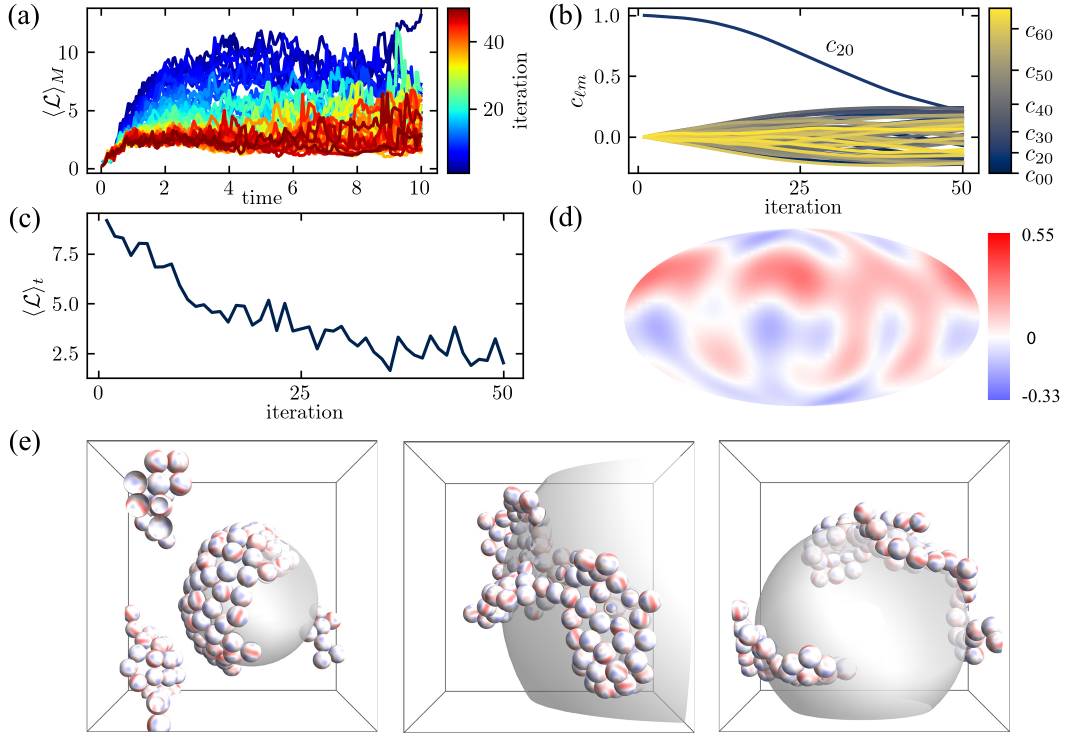}
\caption{Optimization demonstration for a quadrupolar patchy system and target cluster curvature radius $R_t=2.5$. (a) Replica averaged loss values $\langle \mathcal{L}\rangle_M$ in dependence to MD time for all optimization iterations. (b) Change in patch coefficients $c_{\ell m}$ during optimization. (c) Time averaged loss $\langle \mathcal{L}_t\rangle_M$ for the training configuration set. (d) Patch distribution on a particle at iteration $i=50$, plotted in the Mollweide projection. (e) Examples of cluster structure at final MD time for iteration 50: cluster with (mostly) consistent curvature and $R=3.42$, typical cluster with regions of differing curvature and $R=10.23$, and elongated cluster that leads to wrong curvature fit at $R=5.38$.}
\label{fig:quad-opt}
\end{figure*}

Optimization results are shown in Fig.~\ref{fig:quad-opt}. Interaction parameters were set to $\epsilon=30$, $\epsilon_M=18$, $\alpha=1$, $q_0=1.4$ and the energy units are set with $kT\equiv 1$. 
Optimization successfully finds a patch arrangement that breaks the north-south symmetry (as defined by the equatorial placement of the quadrupolar tensor), with attractive and repulsive patches concentrated on different hemispheres. This division is more pronounced compared to the results from the main paper because in the quadrupolar case, patchy interaction needs to acquire a dipolar contribution that is partially already present in the alignment interaction. Nevertheless, the inferred patch distribution is not sufficient to form consistently curved clusters. Quadrupolar interaction itself is symmetric to up-down particle flips which can lead to bindings with ``wrong'' orientations, despite patchy interaction breaking this symmetry. Such bonds particularly often form between larger clusters that are harder to reorient. While some of the final configurations show clusters that approach the target curved structure, these usually only consist of a part of all particles in the box. Larger clusters are irregularly shaped and can be very elongated, which also disturbs the cluster reconstruction algorithm for PBCs, leading to erroneous curve fits. All these effects prevent the optimization algorithm to further improve final cluster shapes and the loss function value stays high -- more than one order or magnitude larger compared to alignment interactions presented in the main text.

\bibliography{references}